\documentclass[a4paper,11pt]{article}
\pdfoutput=1 

\usepackage{jcappub} 

\usepackage[T1]{fontenc} 
\usepackage{amsmath}
\usepackage{multirow}
\usepackage{ulem}
\title{\boldmath Forecasts on CMB lensing observations with AliCPT-1}

\author[a]{Jinyi Liu,}
\author[b,c]{Zeyang Sun,}
\author[a]{Jiakang Han,}
\author[e]{Julien Carron,}
\author[f]{Jacques Delabrouille,}
\author[g]{Siyu Li,}
\author[g]{Yang Liu,}
\author[g]{Jing Jin,}
\author[h,i]{Shamik Ghosh,}
\author[j]{Bin Yue,}
\author[b,c,d]{Pengjie Zhang,}
\author[h]{Chang Feng,}
\author[m,n]{Zhi-Qi Huang,}
\author[k,g]{Hao Liu,}
\author[g]{Yi-Wen Wu,}
\author[m,n,o]{Le Zhang,}
\author[g]{Zi-Rui Zhang,}
\author[h,i]{Wen Zhao,}
\author[a,1]{Bin Hu,\note{Corresponding author.}}
\author[g,1]{Hong Li,}
\author[p]{Xinmin Zhang}

\affiliation[a]{Department of Astronomy, Beijing Normal University, Beijing 100875, People's Republic of China}
\affiliation[b]{Department of Astronomy, School of Physics and Astronomy, Shanghai Jiao Tong University, Shanghai, 200240, People's Republic of China}
\affiliation[c]{Shanghai Key Laboratory for Particle Physics and Cosmology, Shanghai, 200240, People's Republic of China}
\affiliation[d]{Division of Astronomy and Astrophysics, Tsung-Dao Lee Institute, Shanghai Jiao Tong University, Shanghai, 200240, People's Republic of China}
\affiliation[e]{Universit\'e de Gen\`eve, D\'epartement de Physique Th\'eorique et CAP, 24 Quai Ansermet, CH-1211 Gen\`eve 4, Switzerland}
\affiliation[f]{CNRS-UCB International Research Laboratory, Centre Pierre Bin\'etruy, IRL2007, CPB-IN2P3, Berkeley, USA}
\affiliation[g]{Key Laboratory of Particle Astrophysics, Institute of High Energy Physics, Chinese Academy of Sciences, 19B Yuquan Road, Beijing 100049, People’s Republic of China}
\affiliation[h]{CAS Key Laboratory for Researches in Galaxies and Cosmology, Department of Astronomy, University of Science and Technology of China, Chinese Academy of Sciences, Hefei, Anhui 230026, People's Republic of China}
\affiliation[i]{School of Astronomy and Space Science, University of Science and Technology of China, Hefei 230026, People's Republic of China}
\affiliation[j]{National Astronomical Observatories, Chinese Academy of Sciences, Beijing, 100101, People's Republic of China}
\affiliation[k]{School of Physics and optoelectronics engineering, Anhui University, 111 Jiulong Road, Hefei, People's Republic of China}
\affiliation[m]{School of Physics and Astronomy, Sun Yat-sen University, 2 Daxue Road, Tangjia, Zhuhai, 519082, People's Republic of China}
\affiliation[n]{CSST Science Center for the Guangdong-Hong Kong-Macau Greater Bay Area, Zhuhai 519082, People's Republic of China}
\affiliation[o]{Peng Cheng Laboratory, No.2, Xingke 1st Street, Shenzhen 518000, People's Republic of China}
\affiliation[p]{Theoretical devision, Institute of High Energy Physics, Chinese Academy of Sciences, 19B Yuquan Road, Beijing 100049, People’s Republic of China}

\emailAdd{bhu@bnu.edu.cn}
\emailAdd{hongli@ihep.ac.cn}

\abstract{AliCPT-1 is the first Chinese CMB experiment aiming for high precision measurement of Cosmic Microwave Background B-mode polarization. The telescope, currently under deployment in Tibet, will observe in two frequency bands centered at $90$ and $150$\,GHz.
We forecast the CMB lensing reconstruction, lensing-galaxy as well as lensing-CIB (Cosmic Infrared Background) cross correlation signal-to-noise ratio (SNR) for AliCPT-1. 
We consider two stages with different integrated observation time, namely ``$\texttt{4 module*yr}$'' (first stage) and ``$\texttt{48 module*yr}$'' (final stage). For lensing reconstruction, we use three different quadratic estimators, namely temperature-only, polarization-only and minimum-variance estimators, using curved sky geometry. We take into account the impact of inhomogeneous hit counts as well as of the mean-field bias due to incomplete sky coverage. In the first stage, our results show that the $150$ GHz channel is able to measure the lensing signal at $15\sigma$ significance with the minimum-variance estimator. In the final stage, the measurement significance will increase to $31\sigma$. 
We also combine the two frequency data in the harmonic domain to optimize the SNR. 
Our result show that the coadding procedure can significantly reduce the reconstruction bias in the multiple range $\ell>800$.
Thanks to the high quality of the polarization data in the final stage of AliCPT-1, the EB estimator will dominate the lensing reconstruction in this stage. 
We also estimate the SNR of cross-correlations between AliCPT-1 CMB lensing and other tracers of the large scale structure of the universe. For its cross-correlation with DESI galaxies/quasars, we report the cross-correlation SNR = $10\sim20$ for the 4 redshift bins at $0.05<z<2.1$. In the first stage, the total SNR is about $32$. In the final stage, the lensing-galaxy cross-correlation can reach SNR=$52$. 
For lensing-CIB cross-correlation, in the first stage, the cross-correlation between AliCPT-1 lensing and Planck CIB $353,545,857$ GHz channels are about SNR=$18,19,23$ respectively. In the final stage, the cross-correlation can reach SNR=$25,33$ and $42$. Due to the strong correlations between frequency bands, the total lensing-CIB cross-correlation by combining the three frequencies in Planck CIBs are SNR=$23$ and $43$ for the AliCPT-1 first and final stage, respectively.}

\begin{document}
\maketitle
\flushbottom

\section{Introduction to CMB Lensing}
\label{sec:intro}


When the primary CMB photons approach to us from the last scattering surface, they are deflected by the intervening large scale structures which distort the observed pattern of CMB anisotropies. This effect is called CMB lensing \cite{1987A&A...184....1B}. The characteristic deflection angle is about $2$ arcmin. 
Its coherence angular scale is about a few degrees, roughly corresponding to the scale of the peak of the matter power spectrum at $1\lesssim z\lesssim 4$, where the dominant lensing contribution arises.
On the one hand, CMB lensing carries fruitful information of the underlying matter distribution; on the other hand, it does also convert some portion of primary E-mode 
(even parity)
polarization into B-mode 
(odd parity)
polarization, generating lensing B-modes \cite{Kamionkowski:1996ks,Kamionkowski:1996zd,Zaldarriaga:1998ar} which contaminate the measurement of the primordial gravitational wave. Hence, CMB lensing is a useful probe of large scale structure, which can be used to explore the nature of dark energy, dark matter and neutrinos. 
However, for primordial B-mode detection, the lensing B-modes are an unavoidable intrinsic source of noise, with typical amplitude $5~\mu{\rm K}\cdot{\rm arcmin}$ in B-mode map. We recommend references \cite{Lewis:2006fu,Hanson:2009kr} for a review. 

Since the gravitational interaction neither creates nor destroys photons, the total number of CMB photons is conserved during the lensing process. Lensing just remaps the original spatial distributions of CMB photons from the direction $\hat{\textbf{\textit{n}}}$ into a new direction $\hat{\textbf{\textit{n}}}+\textbf{\textit{d}}(\hat{\textbf{\textit{n}}})$, where $\textbf{\textit{d}}(\hat{\textbf{\textit{n}}})$ is called deflection angle with characteristic amplitude $2$ arcmin. Up to the leading order, the deflection angle can be expressed in terms of the gradient of the lensing potential $\textbf{\textit{d}}(\hat{\textbf{\textit{n}}})\simeq\nabla\phi(\hat{\textbf{\textit{n}}})$.
The CMB lensing potential field is defined as 
\begin{equation}\label{eq:potential}
    \phi(\hat{\textbf{\textit{n}}})=-2\int_{0}^{\chi_{\ast}}d\chi \frac{\chi_{\ast}-\chi}{\chi\chi_{\ast}}\Psi(\chi\hat{\textbf{\textit{n}}},\eta_0-\chi)\;,
\end{equation}
where $\chi$ is the conformal distance and $\chi_{\ast}\simeq14~{\rm Gpc}$ the conformal distance between the present and the CMB last scattering surface. $\Psi(\chi\hat{\textbf{\textit{n}}},\eta_0-\chi)$ is the Weyl potential at conformal distance $\chi$ along the direction $\hat{\textbf{\textit{n}}}$ at conformal time $\eta$ (the conformal time today is denoted as $\eta_0$). Here, we explicitly assume flat spatial geometry. 
For the temperature field, we have
\begin{eqnarray}
\label{eq:expansion}
    \tilde{T}(\hat{\textbf{\textit{n}}}) &=& T(\hat{\textbf{\textit{n}}}+\textbf{\textit{d}}(\bold{\hat{n}}))\nonumber\;,\\
    &\simeq& T(\hat{\textbf{\textit{n}}})+ \sum_i\nabla^i\phi(\hat{\textbf{\textit{n}}})\nabla_i T(\hat{\textbf{\textit{n}}})+\mathcal{O}(\phi^2)\;.
\end{eqnarray}
For polarization fields, we have 
\begin{eqnarray}
    \left[\tilde{Q}\pm i\tilde{U}\right](\hat{\textbf{\textit{n}}}) &=& \left[Q\pm iU\right](\hat{\textbf{\textit{n}}}+\textbf{\textit{d}}(\hat{\textbf{\textit{n}}}))\nonumber\;,\\
    &\simeq&\left[Q\pm iU\right](\hat{\textbf{\textit{n}}})+\sum_i\nabla^i\phi(\hat{\textbf{\textit{n}}})\nabla_i\left[Q\pm iU\right](\hat{\textbf{\textit{n}}})+\mathcal{O}(\phi^2)\;,
\end{eqnarray}
where the quantities with and without tilde correspond to the lensed and primary CMB, respectively. 

The first CMB lensing signal was detected via the cross-correlations between WMAP\,1-year temperature data and Sloan Digital Sky Survey (SDSS) Luminous Red Galaxies \cite{Hirata:2004rp}. Subsequently, several  teams tried to reconstruct the lensing signal using other large-scale structure tracers \cite{Smith:2007rg,Hirata:2008cb,Feng:2012uf}. 
At low signal to noise ratio, the lensing signal has been detected with WMAP\,7-year temperature data internally \cite{Smidt:2010by,Feng:2011jx}. After that, series of ground-based CMB experiments have measured the lensing signal with higher signal to noise ratio. The Atacama Cosmology Telescope (ACT) collaboration in 2011 firstly presented a $4.0\sigma$ detection of CMB lensing signal internally with their temperature map \cite{Das:2011ak}, updated in 2013 with a $4.6\sigma$ detection \cite{Das:2013zf}. 
The South Pole Telescope (SPT) collaboration reported their
first lensing power spectrum reconstruction result in 2012 \cite{vanEngelen:2012va}. Later on, the first detection of lensing B-modes was made by SPT in 2013 by using the cross-correlation between maps of CMB polarization and sub-mm maps of galaxies from {\it Herschel}-SPIRE \cite{Hanson:2013hsb}. By using $500{\rm deg}^2$ of SPTpol data from $95$ GHz and $150$ GHz channels, the collaboration measured the BB power spectrum in the multipole range $25<\ell<2301$ with $18.1\sigma$ detection, among which lensing B-modes were detected with $8.7\sigma$ significance \cite{Sayre:2019dic}. 
The POLARBEAR collaboration detects the lensing signal at $4.2\sigma$ confidence level from $30{\rm deg}^2$ polarization map \cite{Ade:2013gez} and at $4.0\sigma$ ($2.3\sigma$ for lensing B-modes) confidence level from cross-correlation with the {\it Herschel} cosmic infrared background \cite{Ade:2013hjl}. Thanks to the excellent sensitivity ($\sim ~3~\mu{\rm K}\cdot{\rm arcmin}$) BICEP2 $\&$ Keck array are able to measure the lensing signal at $5.8\sigma$ level, with a modest angular resolution ($\sim 0.5^{\circ}$) \cite{Array:2016afx}.

The first reconstructed lensing potential map on the nearly full sky is obtained by the Planck collaboration in 2013 \cite{Ade:2013tyw} with $25\sigma$ significance. This lensing map is reconstructed from the $15$ months temperature data alone. In 2015, the Planck collaboration updated this reconstruction by adding another $15$ months of temperature and $30$ months full-mission of polarization data \cite{Ade:2015zua}. These additional data help increase the lensing reconstruction significance up to $40\sigma$ level. The final full-mission analysis in 2018 uses essentially the same data as 2015, but improves the foreground masking in the simulations. It helps increase the significance of the detection of lensing in the polarization maps alone from $5\sigma$ to $9\sigma$ \cite{Aghanim:2018oex}. Furthermore, the collaboration also demonstrates the delensing technique with the final full-mission data. A decrease in power of the B-mode polarization after delensing is detected at $9\sigma$. Up to the knowledge of the authors, this $10.1\sigma$ detection from SPTpol \cite{Wu:2019hek} and $9\sigma$ detection from Planck 2018 \cite{Aghanim:2018oex} represents the state-of-the-art measurement of the lensing signal from polarization data. We can treat this number as the benchmark for lensing reconstruction from polarization data. From the rest of this paper, one can see that the polarization data from AliCPT-1 has the capability of improving this number significantly. This makes one of the extraordinary science cases for AliCPT-1 project.    

CMB lensing is also physically correlated with LSS tracers such as galaxy and galaxy cluster distribution, and cosmic shear. On one hand, these cross-correlations are immune to certain systematics in the auto-correlation, such as the additive errors in both CMB lensing and cosmic shear. On the other hand, they provide essential information, such as the redshift information, to improve the cosmological applications of CMB lensing. Cross-correlation of CMB lensing with galaxies \cite{Darwish:2020fwf, Kitanidis:2020xno, Marques:2020dsb, Krolewski:2021yqy, White:2021yvw}, galaxy groups and clusters \cite{Geach:2017crt, DES:2017fyz, ACT:2020izl, Sun:2021rhp} have already been detected. Since the AliCPT sky area is fully covered by the DESI footprint \cite{DESI:2018ymu}, AliCPT CMB lensing-DESI galaxy cross-correlation is a natural way of enhancing the science return of AliCPT. For this reason, we estimate the cross-correlation signal between AliCPT and DESI. We find that the total SNR is $\sim 32$, which will provide useful constraint on the structure growth rate.

The cosmic infrared background (CIB) carries the integrated history of star formation between redshift $1\leq z\leq 3$, which highly overlaps with the CMB lensing signal. The first detection of this cross correlation was reported by Planck collaboration at 2013 \cite{Planck:2013qqi} by correlating CMB lensing with CIB measured in $353$, $545$ and $857$ GHz channels. As an external tracer, CIB can also be used to delens. 
Via the delensing technique, the Planck 2018 result \cite{Aghanim:2018oex} reported the primary CMB peak sharpening at $11\sigma$ from the lensing reconstruction alone and $15\sigma$ on further combining with CIB.

\section{AliCPT-1 mock data setup}
\label{sec:mock}

AliCPT-1 is the first Chinese CMB experiment. Its main science goal is to constrain the primordial gravitational wave signal with high precision. The telescope has two frequencies, namely $90$ and $150$ GHz. The full focal plane can accomodate $19$ modules containing $1704$ TES detectors each, which are equally distributed between two frequencies.      

In this section, we introduce the mock data for each single frequency band.
According to the scanning strategy, we get the number of hits per pixel. The noise level in each frequency band is proportional to the ratio of ``NET'' (Noise Equivalent Temperature) and square root of the hit counts, i.e.  
\begin{equation}
\label{eq:noise}
    {\rm noise} = {\rm NET}/\sqrt{\rm hits}\;.
\end{equation}
In our simulations, we assume there is no correlation between AliCPT temperature and polarization noise. The NET for polarization is a factor $\sqrt{2}$ higher than that of temperature since the observation time is spread between two polarization signals, the Q and U Stokes parameters.
Because the number of detectors in each frequencies are evenly distributed in the modules, the noise spatial distribution in the two frequencies are almost the same up to an overall normalization factor. Here, we explicitly assume the noise to be uncorrelated between different pixels. 

We consider two different accumulated module numbers for the nominal and final AliCPT-1 experiment. The first stage uses $4$ modules observing for $1$ year, is dubbed hereafter the ``$\texttt{4 module*yr}$'' stage. 
The preliminary roadmap for the final mission is: the first year observes with $4$ modules; the second year adds $6$ modules; the third year adds another $5$ modules; and finally the last $4$ modules are added in the focal plane for the last year\footnote{The focal plane can assemble $19$ modules in total.}. Hence, the total integrated observation time will be ``$\texttt{48 module*yr}$'' after four years of observation. In this paper, we forecast the lensing reconstruction, lensing-galaxy and lensing-CIB cross-correlations based on these two integrated observation time. We note that in practice, the actual time of observation is less, because of bad weather and various data cuts. This is taken into account in the simulations. 

\begin{figure}
\centering
\includegraphics[width=0.9\linewidth]{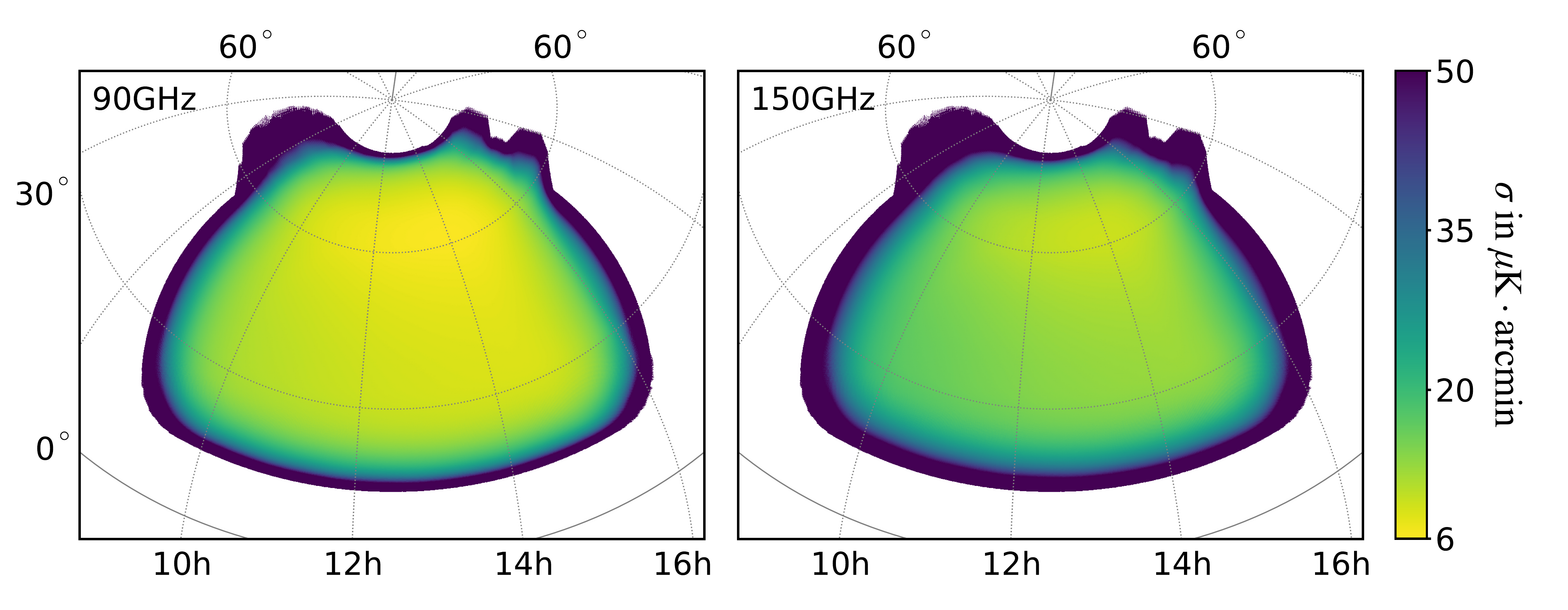}
\caption{Noise variance map for temperature mode in $90$ and $150$ GHz under the ``$\texttt{4 module*yr}$'' stage. The noise variance for polarization mode can be obtained by multiplying a factor $\sqrt{2}$.}
\label{fig:noise_var}
\end{figure}

In Figure \ref{fig:noise_var}, we show the noise variance maps for ``$\texttt{4 module*yr}$'' stage. 
From now on, we simply consider the statistical noise, the variance in the ``$\texttt{48 module*yr}$'' stage can be obtained by rescaling the ``$\texttt{4 module*yr}$'' one with an overall normalization factor $1/\sqrt{12}$.  
The left and right panels of Figure \ref{fig:noise_var} are the variance map in the $90$ and $150$ GHz channels, respectively. These are obtained by adopting the present ``deep patch'' scanning strategy\footnote{We refer the details of the ``deep patch'' scanning strategy in the simulation paper of this series.}, which covers $14\%$ sky area. The harmonic mean of the noise variance in ``$\texttt{4 module*yr}$'' case are about $11~\mu{\rm K}\cdot{\rm arcmin}$ for $90$ GHz and $17~\mu{\rm K}\cdot{\rm arcmin}$ for $150$ GHz.   

The mock data sets in each frequency channel contain $301$ sets of simulations, among which $240$ sets are used for covariance estimation, $60$ sets are used for mean field subtraction and $1$ set represents the ``real'' data. For each of these $301$ data sets, we include the unlensed primary CMB maps ($T,E,B$), a lensing potential map ($\phi$) to lens the primordial maps, and a noise realization according to the noise variance map shown in Figure~\ref{fig:noise_var}. All of the data sets are generated following the same procedure but with independent Gaussian random seeds. The CMB maps, lensing potential maps as well as the noise maps among different data sets are independent. 
The primary CMB and lensing potential maps are generated via \texttt{CAMB} \cite{Lewis:1999bs} and \texttt{HEALPix} \cite{Gorski:2004by} with resolution parameter ${\rm N_{side}}=1024$.
The lensed CMB maps are obtained by combining the unlensed primary CMB with the lensing potential maps via \texttt{lenspyx} code\footnote{https://github.com/carronj/lenspyx} \cite{2020ascl.soft10010C}. 
Then, we convolve the beam function ($B_{\ell}$) as well as the \texttt{HEALPix} transfer function ($H_{\ell}$) with the lensed CMB maps in the pixel domain. The full width at half maximum (FWHM) of the beam function are $19$ arcmin for the $90$ GHz channel, and $11$ arcmin for the $150$ GHz channel. Finally, we add the noise in the pixel domain directly on top of the lensed CMB signals convoved by the Gaussian beams.  

In this paper, we adopt the curved sky quadratic estimators developed by Okamoto and Hu in 2003 \cite{Okamoto:2003zw}.
This estimator was first introduced by Hu and Okamoto in 2001 \cite{Hu:2001kj} with a flat sky approximation. 
Since AliCPT-1 has $14\%$ sky coverage, to handle the lower multipoles better, we decide to use the mathematically more complicated curved sky quadratic estimator, in the slightly modified version called `SQE' by Ref.~\cite{Maniyar:2021msb}.

\section{Lensing reconstruction}
\label{sec:recon}
In this section, we present the quadratic lensing reconstruction methodology and results. 

\subsection{Method}
\label{subsec:method}

Lensing induces correlations between different multipoles. The off-diagonal term of the covariance matrix of the CMB fields reads 
\begin{equation}
\label{eq:off-diag}
  \langle X_{\ell_1m_1}Z_{\ell_2m_2}\rangle = \sum\limits_{LM}(-1)^M
  \begin{pmatrix}
    \ell_1,& \ell_2,& L \\
    m_1,& m_2,& -M\\
  \end{pmatrix}
  \mathcal{W}^{XZ}_{\ell_1\ell_2L}\phi_{LM}\;,
\end{equation}
where the fields $X_{\ell m}, Z_{\ell m}\in \{T_{\ell m}, E_{\ell m}, B_{\ell m}\}$. The expression of the covariance response function $\mathcal{W}^{XZ}_{\ell_1\ell_2L}$ for all the possible field combinations can be found in Okamoto and Hu \cite{Okamoto:2003zw}. The big parenthesis denotes the Wigner 3-$j$ symbol due to the coupling of different angular momentum modes. $\phi_{LM}$ is the lensing potential spherical harmonics, namely our reconstruction goal. Owing to the nearly-Gaussian nature of the primary CMB signal at recombination, there is no correlation between different multipoles of the primordial components. Hence, the lensing contribution appears in the leading order in the off-diagonal terms. The basic idea of the quadratic estimator is to utilize these induced correlations in an (almost) optimal and unbiased way. In what follows, we will list a few essential steps of our reconstruction recipe. 

\subsubsection{Filtering}

In this paper, we follow the Planck 2018 lensing paper\footnote{Hereafter, we will refer to this paper as PL18.} \cite{Aghanim:2018oex} formalism, which is close to the original Hu-Okamoto formalism \cite{Hu:2001kj,Okamoto:2003zw}. The basic idea is to rewrite the lensing estimator in terms of pairs of filtered maps \cite{Carron:2017mqf}. One leg of the pair is the inverse-variance filtered CMB fields; the other leg is the Wiener-filtered CMB fields. The inverse-variance filter operation is defined as
\begin{equation}
    \label{eq:ivf_field}
    \bar X(\hat n)=\left[\mathcal{B}^{\dagger}{\rm CoV}^{-1}X^{\rm dat}\right](\hat n)\;,
\end{equation}
where matrix $\mathcal{B}$ accounts for the real-space operations of beam and pixel convolution. 
$X^{\rm dat}(\hat n)\in{T,E,B}$ are the simulated/observed data.
${\rm CoV}$ is the pixel-space covariance defined as ${\rm CoV}=\mathcal{T}\mathcal{C}^{\rm fid}\mathcal{T}^{\dagger}+\mathcal{N}$, 
with $\mathcal{C}^{\rm fid}$ denotes the fiducial CMB spectra, $\mathcal{N}$ denotes for the pixel-space noise covariance matrix, which is approximated as diagonal, and $\mathcal{T=BY}$ denotes the complete transfer function from multipoles to the pixelized sky. The matrix $\mathcal{Y}$ contains the appropriate (spin-weighted) spherical harmonic functions to map from multipoles to the sky. 
The Wiener-filtered CMB fields are defined as
\begin{equation}
\label{eq:Wienerf}
  \begin{pmatrix}
    T^{\rm WF} &\\
    E^{\rm WF} &\\ 
    B^{\rm WF} &
  \end{pmatrix}
  = \mathcal{C}^{\rm fid}\mathcal{T}^{\dagger}{\rm CoV}^{-1}
  \begin{pmatrix}
     & T^{\rm dat} &\\
     & {}_{+2}P^{\rm dat} &\\ 
     & {}_{-2}P^{\rm dat} &
  \end{pmatrix}\;.
\end{equation}
Notice that here, we filter the temperature and polarization data independently, which means that we neglect the TE correlation in the covariance matrix of Eq. (\ref{eq:Wienerf}). As demonstrated in PL18, neglecting $C^{TE}_{\ell}$ in the Wiener filter process will cause about $3\%$ noise increase for the multipoles $L<400$, and less in our case. The inversion of the covariance matrix is computed via a conjugate-gradient inversion method with a multi-grid preconditioner~\cite{Smith:2007rg}.

\subsubsection{Quadratic estimator}

Then, one can construct a (yet unnormalized) lensing deflection angle (spin-1 field) estimate based on the inverse-variance filtered and Wiener filtered fields 
\begin{equation}
    \label{eq:destimator}
    {}_{1}\hat d(\hat{{\bold n}})=-\sum\limits_{s=0,\pm 2}{}_{-s}\bar X(\hat{{\bold n}})\left[\eth_sX^{\rm WF}\right](\hat{{\bold n}})\;,
\end{equation}
where $\eth$ is the spin-raising operator, and the pre-subscript $s$ on the field indicates  spin. The explicit expression of the spin-raising operator can be found in PL18. The deflection angle ${}_{1}\hat d(\hat{{\bold n}})$ can be decomposed into its gradient-like (g) and curl-like (c) component via the spin-1 harmonic transformation. The expected curl-like component (corresponding to field rotation\cite{Hirata:2003ka,Lewis:2006fu}) is safely negligible for our purposes, while the gradient term directly traces the lensing potential $\phi$
\begin{equation}
\label{eq:gnc}
    {}_{\pm 1}\hat d(\hat{{\bold n}})=\mp \sum_{LM}\left(\frac{\hat g_{LM} \pm i\hat c_{LM}}{\sqrt{L(L+1)}}\right){}_{\pm 1}Y_{LM}(\hat{{\bold n}})\;.
\end{equation}
Following PL18, we calculate three estimators, namely temperature-only (T-only) ($s=0$), polarization-only ($s=\pm 2$) (P-only), and minimum-variance (MV) ($s=0,\pm 2$),
rather than the traditional full set TT, TE, TB, EE and EB of Hu-Okamoto estimators.
In the T-only and P-only estimators, we analyse the temperature and polarization data independently and do not need to include the TE cross correlation. For MV estimator, we do include TE correlation in the fiducial spectra matrix $\mathcal{C}^{\rm fid}$ in Eq. (\ref{eq:Wienerf}) (resulting in the generalized MV estimator of Ref.~\cite{Maniyar:2021msb}).  

\subsubsection{Mean-fields subtraction}

The presence of a mask and foreground residuals introduce extra statistical anisotropies in the absence of lensing signal. Hence, they will bias the estimation of lensing potential. Since these effects are hard to model analytically, we calculate their contributions via Monte Carlo simulations. Namely, in each of the simulation sets we vary simultaneously multiple ingredients, such as the primary CMB, lensing potential and instrumental noise. Then, we calculate the average value (mean-fields) of the quadratic estimator. We believe this average will be the most faithful representation of the contribution from the extra statistical anisotropy source. Hence, we can subtract it from the original estimator. This operation is called mean-field subtraction. Our lensing potential estimate becomes
\begin{equation}
    \label{eq:meanfield}
    \hat\phi_{LM} = \frac{1}{\mathcal{R}^{\phi}_L}(\hat g_{LM}-\langle\hat g_{LM}\rangle_{\rm MC})\;,
\end{equation}
where $\mathcal{R}^{\phi}_L$ is the isotropic normalization, which is called response function. It is calculated analytically \cite{Ade:2015zua}
\begin{equation}
\label{eq:RL}
    \mathcal{R}^{\phi}_L=\frac{1}{2(2L+1)}\sum_{\ell_1,\ell_2}\mathcal{W}^{XZ}_{\ell_1\ell_2L}M^{XZ}_{\ell_1\ell_2L}F^X_{\ell_1}F^Z_{\ell_2}\;,
\end{equation}
where $\mathcal{W}^{XZ}_{\ell_1\ell_2L}$ is the covaraince off-diagonal response function defined in Eq. (\ref{eq:off-diag}), $M^{XZ}_{\ell_1\ell_2L}$ the XZ-estimator weighting function and $F^X_{\ell}$ is the isotropic Wiener filter
\begin{equation}
F^X_\ell=\frac{C^{XX}_\ell}{C^{XX}_\ell+N^{XX}_\ell}\;.
\end{equation}

\subsubsection{Lensing potential power spectrum and statistical bias}

\begin{figure}
\centering 
\includegraphics[width=.9 \textwidth]{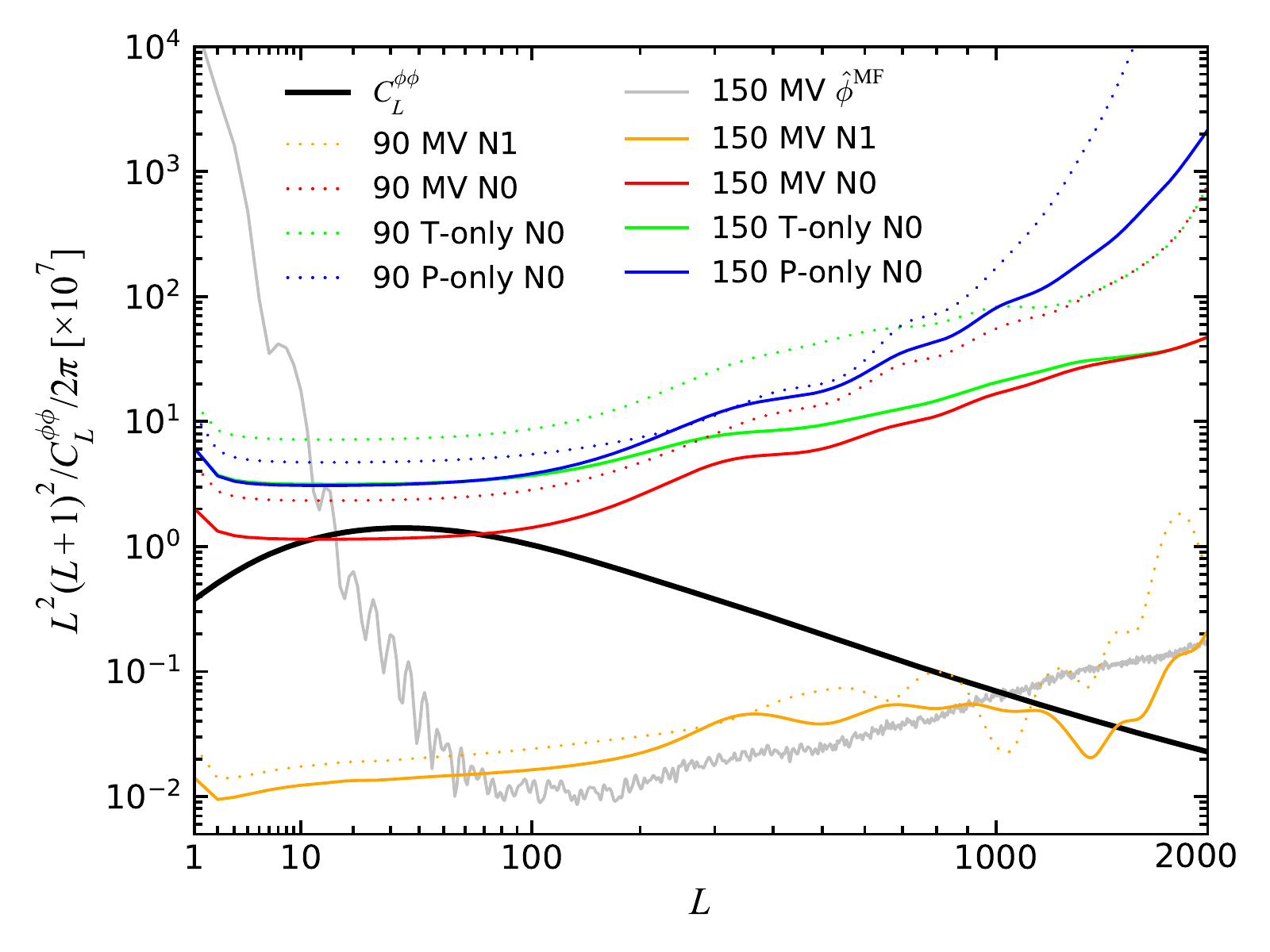}
\caption{\label{fig:cl_noise} Noise power spectra in lensing reconstruction under the ``$\texttt{4 module*yr}$'' stage. The green, blue and red curves are the $N^{(0)}$ bias for T-only, P-only and MV estimators. Orange curves are the $N^{(1)}$ biases in the MV estimators. Solid and dotted curves are for $150$ and $90$ GHz, respectively. The solid grey curve denotes for the mean-field bias in the $150$ GHz MV estimator. The bold solid black curve denotes for the theoretical lensing potential spectrum.}
\end{figure}

The raw power spectrum of the estimated lensing potential reads
\begin{equation}
    \label{eq:phihatpow}
    \hat{C}_L^{\hat\phi_1\hat\phi_2}=\frac{1}{(2L+1)f_{\rm sky}}\sum_{M=-L}^{L}\hat\phi_{1,LM}\hat\phi_{2,LM}^{\ast}\;,
\end{equation}
where $\hat\phi_{1,LM},\hat\phi_{2,LM}$ are (possibly different) estimates of the lensing potential, Eq. (\ref{eq:meanfield}). The quadratic estimator spectrum contains the sought-after signal, but also unavoidably Gaussian reconstruction noise sourced by the CMB and instrumental noise ($N^{(0)}$ bias), as well as the non-primary couplings of the connected 4-point function \cite{Kesden:2003cc} ($N_1$ bias). After subtracting these biases, we get the final estimated power spectrum
\begin{equation}
    \label{eq:phipow}
    \hat C^{\phi\phi}_L=\hat C_L^{\hat\phi_1\hat\phi_2}-\Delta C_L^{\hat\phi_1\hat\phi_2}|_{\rm RDN0}-\Delta C_L^{\hat\phi_1\hat\phi_2}|_{\rm N1}-\cdots\;,
\end{equation}
where the dots denote possible other contamination biases which are not considered here, such as point sources.  
``RDN0'' means realization-dependent $N^{(0)}$ bias, which is designed to subtract the primary CMB contamination in the most faithful manner.
For further detailed expressions, we refer to the Appendix A of Planck 2015 lensing paper \cite{Ade:2015zua}. 

In Figure \ref{fig:cl_noise}, we show several noise spectra in the reconstruction in the ``$\texttt{4 module*yr}$'' stage. We highlight the MV-N0 noise spectrum in the $150$ GHz channel (red solid curve). One can see that the signal (black bold curve) is higher than this noise curve for each multipole in the range $10\leq L \leq100$. 

\subsubsection{Binning and multiplicative correction and SNR estimation}

After binning the multipoles, the band power of the lensing potential reads
\begin{equation}
    \label{eq:bandpow}
    \hat C^{\phi\phi}_{L_b}=\left(\sum_L\mathcal{B}_b^L\hat C^{\phi\phi}_L\right)\left(\frac{\sum_L\mathcal{B}_b^L C_L^{\phi\phi,{\rm fid}}}{\sum_L\mathcal{B}_b^L\langle\hat C^{\phi\phi}_L\rangle_{\rm MC}}\right)\;,
\end{equation}
where the binning function is defined as
\begin{equation}
    \label{eq:binning}
    \mathcal{B}_b^L=C^{\phi\phi,{\rm fid}}_{L_b}\frac{C_L^{\phi\phi,{\rm fid}}V^{-1}_L}{\sum_{L'}(C_{L'}^{\phi\phi,{\rm fid}})^2V_{L'}^{-1}}\;,\;\;\; L_{\rm min}^b\leq L\leq L^b_{\rm max}\;.
\end{equation}
This binning method is designed to produce the minimum variance with optimal weights, which reads
\begin{equation}
    V_L^{-1}\propto (2L+1)f_{\rm sky}(\mathcal R^{\phi}_L)^2\;.
\end{equation}
$L_b$ is the band power index,
\begin{equation}
    \label{eq:Lb}
    L_b=\frac{\sum_L L\mathcal{B}_b^L}{\sum_{L'}\mathcal{B}_b^{L'}}\;.
\end{equation}
The final ingredient of the power spectrum reconstruction is the second parenthesis in the right hand side of Eq. (\ref{eq:bandpow}), namely ``multiplicative correction''. It corrects for the various isotropic and simplified approximations. The $\langle\hat C^{\phi\phi}_L\rangle_{\rm MC}$ in the denominator is calculated via the much cheaper Monte Carlo $N^{(0)}$ estimation, rather than the RDN0 method.
In this paper, we estimate the signal-to-noise ratio (SNR) via the Fisher matrix method
\begin{equation}
    {\rm SNR}=\sqrt{\sum_{\ell,\ell'}C_{\ell}\mathbb{C}^{-1}_{\ell\ell'}C_{\ell'}}\;,
\end{equation}
where the $C_{\ell}$'s in the numerator are chosen to be the theoretical one instead of the reconstructed one. The latter (reconstructed spectrum) has unavoidable random scatter, which may affect the final SNR prediction. In order to make a stable prediction, here we choose the former one (theoretical spectrum).
$\mathbb{C}_{\ell\ell'}$ is the covariance matrix obtained from 240 simulation sets. 

\subsection{Result}
\label{sec:result}

In this subsection, we summarise our lensing reconstruction forecast results based on the mock data presented in Section \ref{sec:mock} and the methodology reviewed in subsection \ref{subsec:method}. 

\begin{figure}
\centering 
\includegraphics[width=.9 \textwidth]{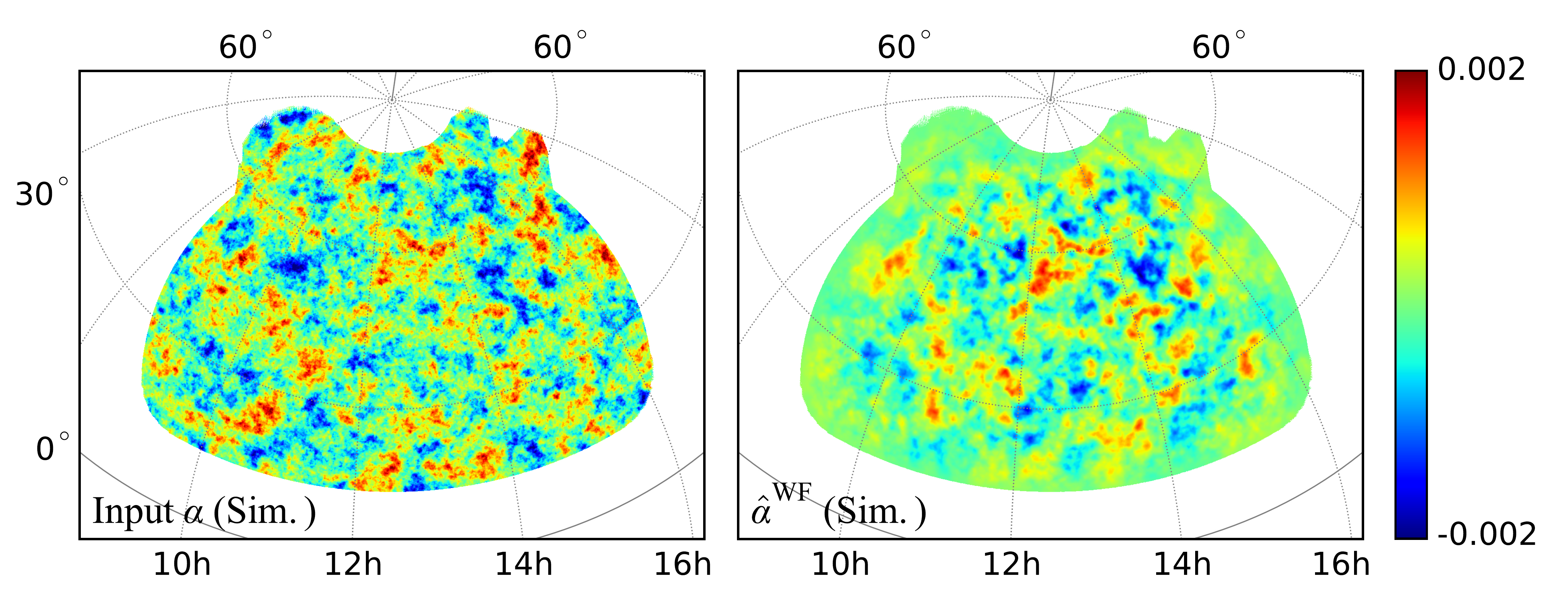}
\caption{\label{fig:phi_map} Input (left) and MV-reconstructed (right) Wiener-filtered deflection angle, $\alpha=|\nabla\phi|$. The reconstruction is done in the $\texttt{4 module*yr}$ case.}
\end{figure}

In Figure \ref{fig:phi_map}, we show the input (left panel) and MV reconstructed (right panel) deflection angle under the $\texttt{4 module*yr}$ scenario. In order to highlight the signal, we show the result after Wiener filtering. Since the noise on the edge increases drastically, the filtered signal on the edges get suppressed significantly. 
Thanks to the long enough integrated observation time, in the center of the field, we do recover the original lensing deflection field large-scale features.  

\begin{table}
\small
\centering
\caption{Lensing reconstruction SNR}
\vspace{2mm}
\label{tab:recsnr}
\begin{tabular}{|c|c|c|c|}
  \hline
  frequency & estimator & $\texttt{4 module*yr}$ & $\texttt{48 module*yr}$ \\
  \hline
    $\multirow{3}{*}{90 GHz}$ & T-only &$3.5$ & $4.8$ \\
                                                     \cline{2-2}
    & P-only & $5.1$ & $21.7$  \\
                                                     \cline{2-2}
    & MV & $9.2$ & $24.3$ \cr 
    \hline
        $\multirow{3}{*}{150 GHz}$ & T-only & $8.3$ & $8.0$ \\
                                                     \cline{2-2}
    & P-only & $6.6$ & $25.4$  \\
                                                     \cline{2-2}
    & MV & $15.4$ & $31.1$ \cr
    \hline
\end{tabular}
\end{table}

\begin{figure}
    \centering
    \includegraphics[width=.9 \textwidth]{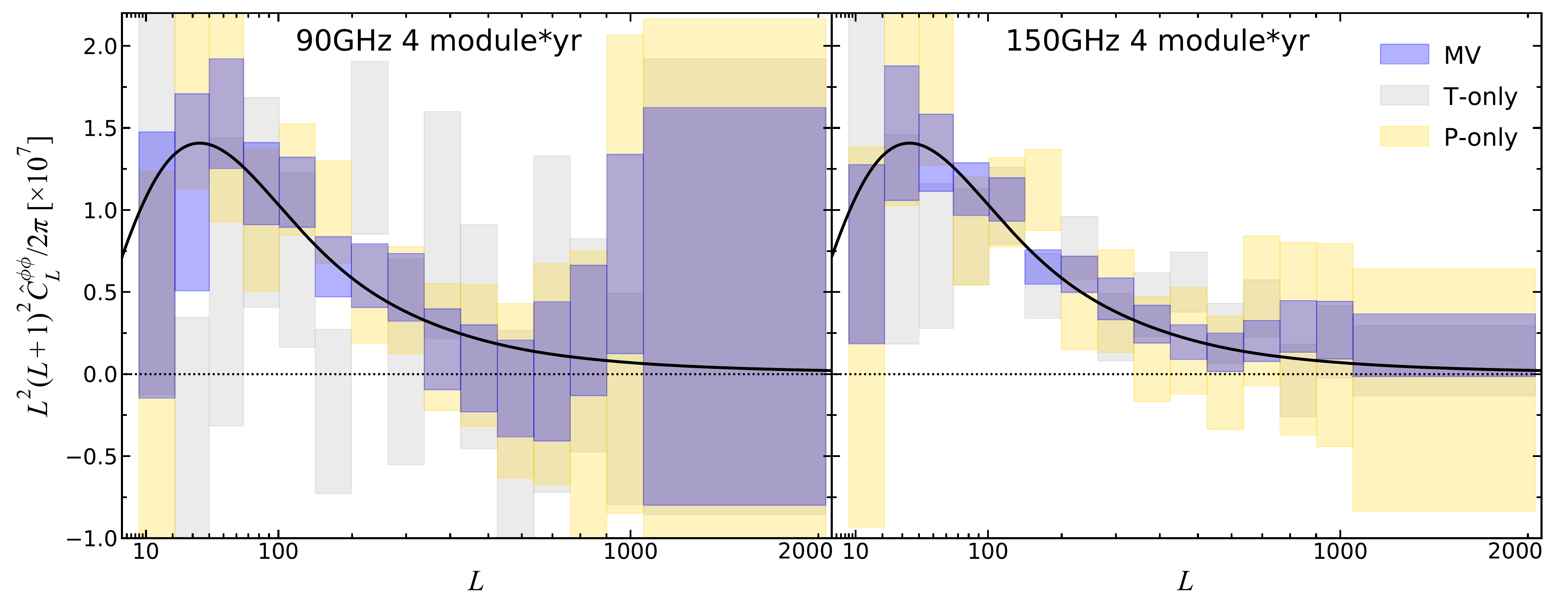}
    \caption{Power spectrum reconstruction results in ``$\texttt{4 module*yr}$'' configuration. In $90$ GHz channel, SNR are $3.5,5.1,9.2$ for T-only, P-only and MV estimators. In $150$ GHz channel, SNR are $8.3,6.6,15.4$ for T-only, P-only and MV estimators.}
    \label{fig:recon_cl_4mod}
\end{figure}

In Figure \ref{fig:recon_cl_4mod}, we show the reconstructed lensing potential spectra in the ``$\texttt{4 module*yr}$'' stage. In the $90$ GHz channel, the SNR are $3.5,5.1,9.2$ for T-only, P-only and MV estimators. In the $150$ GHz channel, the SNR are $8.3,6.6,15.4$ for T-only, P-only and MV estimators. For the MV estimator (purple boxes), we can get SNR higher than unity in each of the multiple bands in the range of $L<400$ ($90$ GHz) and $L <700$ ($150$ GHz). This is the main result of this paper.

\begin{figure}
    \centering
    \includegraphics[width=.9 \textwidth]{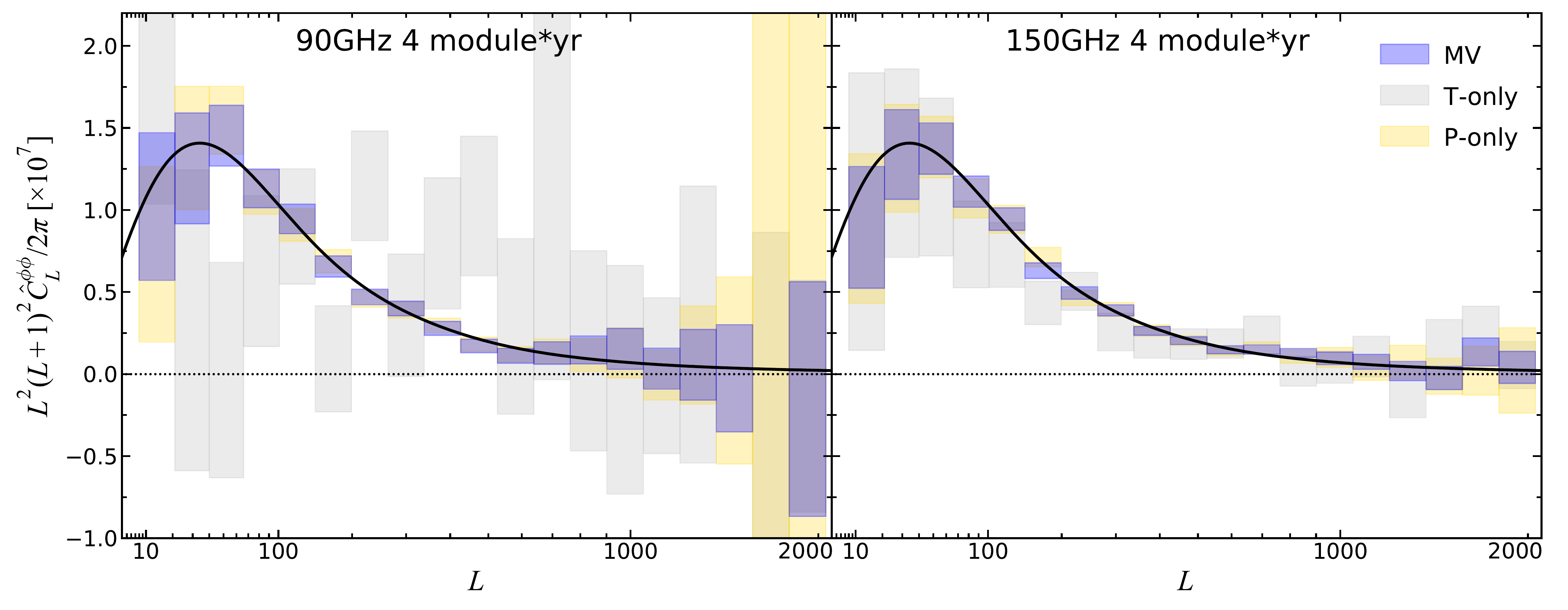}
    \caption{Power spectrum reconstruction results in ``$\texttt{48 module*yr}$'' configuration. In $90$ GHz channel, SNR are $4.8,21.7,24.3$ for T-only, P-only and MV estimators. In $150$ GHz channel, SNR are $8.0,25.4,31.1$ for T-only, P-only and MV estimators.}
    \label{fig:recon_cl_48mod}
\end{figure}

In Figure \ref{fig:recon_cl_48mod}, we show the reconstructed lensing potential spectra under the ``$\texttt{48 module*yr}$'' stage. In the $90$ GHz channel, SNR are $4.8,21.7,24.3$ for T-only, P-only and MV estimators. In the $150$ GHz channel, SNR are $8.0,25.4,31.1$ for T-only, P-only and MV estimators. For the MV estimator (purple boxes), we can get SNR higher than unity in each of the multiple bands in the range $L<900$ ($90$ GHz) and $L<1100$ ($150$ GHz). We summarize the reconstruction SNR in Table \ref{tab:recsnr}. 
We further analyse the contributions from TE and EB estimators. Take the 150 GHz channel as an example, in the ``$\texttt{4 module*yr}$'' stage, the SNR from TE (SNR=6.4) is higher than that of EB (SNR=3.1). Once we accumulate the data and step in the ``$\texttt{48 module*yr}$'' stage, the EB estimator contributes significantly, the corresponding SNR=21.4. In this stage, the TE estimator becomes sub-dominant, SNR=13.8. This result demonstrates the important role of polarization data in the AliCPT lensing analysis. This is another major result of this paper.

\begin{figure}
    \centering
    \includegraphics[width=.9 \textwidth]{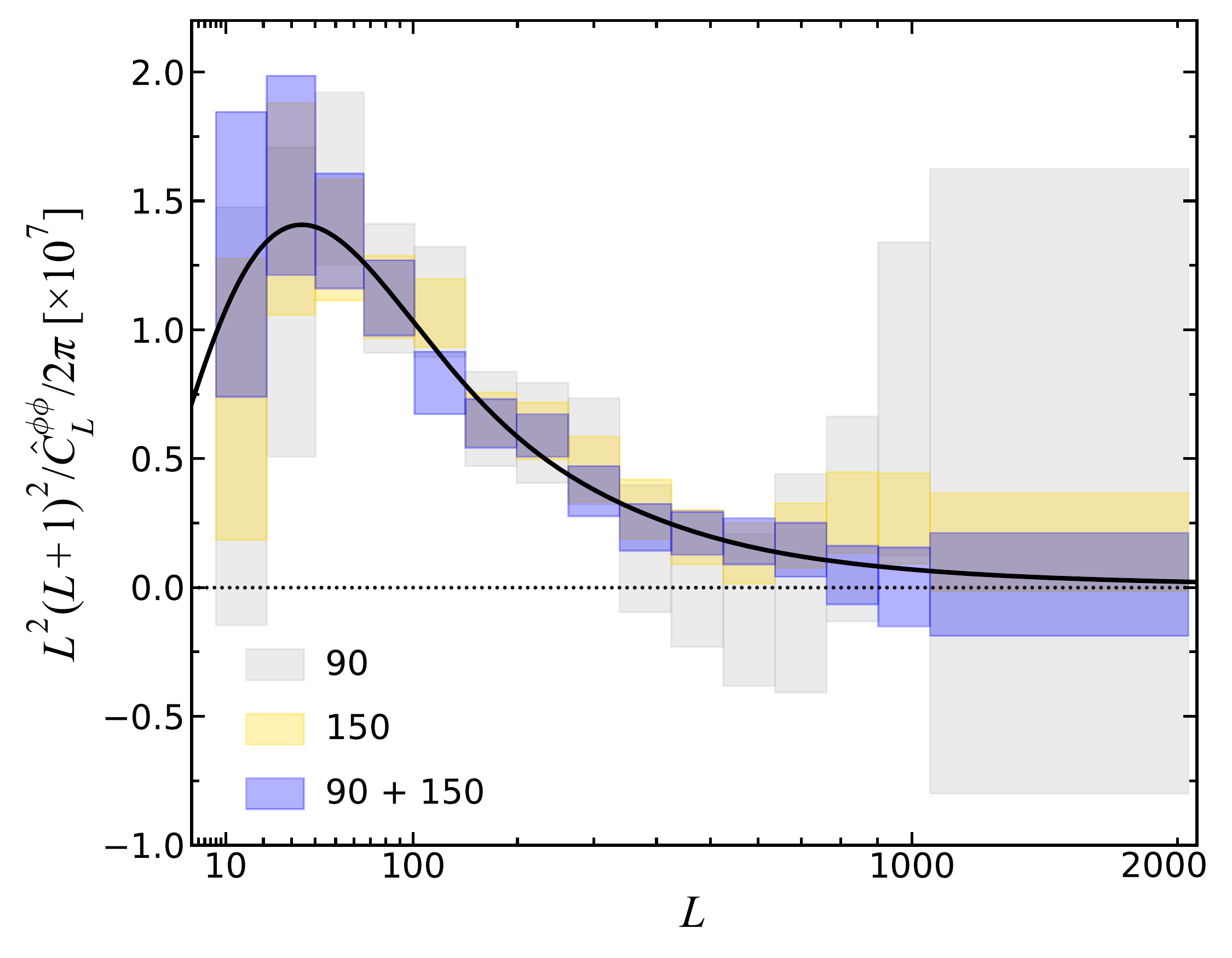}
    \caption{Coadding result in the ``$\texttt{4 module*yr}$'' stage.
    Gray, yellow and purple boxes are the MV estimator reconstructions from $90$GHz, $150$GHz and $90+150$GHz.
    Comparing with the $150$GHz channel, the coadding procedure does not reduce the error bars but remove the systematic bias in the multiple range $\ell>800$.}
    \label{fig:coadding}
\end{figure}

Furthermore, we tested the coadding results by combining $90$ and $150$ GHz maps in the harmonic domain with the inverse minimum-variance weighting. This method is basically the same idea as the ``SMICA weights'' in Planck 2015 and 2018 lensing paper \cite{Ade:2015zua,Aghanim:2018oex}. We show the MV estimator coadding results in the ``$\texttt{4 module*yr}$'' stage in Figure \ref{fig:coadding}. Comparing the yellow ($150$GHz) with the purple ($90+150$GHz) boxes, one can see that the coadding procedure can significantly reduce the systematic bias in the multiple bins above $\ell=800$. As for the error bars, we find 
that the improvement by combining the two frequencies is very limited and can be ignored. This is due to the beam width in $90$ GHz (FWHM=$19$ arcmin) is too large to help increase the total signal. The final coadded signal is nearly all from $150$ GHz, which has $11$ arcmin beam width. Similar results are also found by Planck 2013 lensing paper \cite{Ade:2013tyw}. In their analysis, they reported that the $100$ GHz (FWHM=$10$ arcmin) helps very little compared with $143$ and $217$ GHz, which have $7$ and $5$ arcmin beams, respectively. Hence, in the nominal data analysis, they do not include $100$ GHz channel at all.  

\section{Cross-correlations}
\label{sec:XC}
In this section, we will present the cross-correlation signals which can be detected by AliCPT CMB lensing together with the external large scale structure tracers, such as galaxy number counts and cosmic infrared background. 

\subsection{Cross-correlations with galaxies}
\label{sec:XCgal}
It is convenient to work with the lensing convergence $\kappa$, instead of potential $\phi$, for the lensing-galaxy cross-correlation
\begin{equation}
    \kappa(\boldsymbol{\theta}) = \frac{3\Omega_{m0}H_0^2}{2c^2} \int_0^{\chi_\ast}d\chi \frac{\chi(\chi_\ast-\chi)}{\chi_*}\frac{\delta_m(\chi, \boldsymbol{\theta})}{a} = \int_0^{\chi_\ast}d\chi W^\kappa(\chi)\delta_m(\chi,\boldsymbol{\theta}) \ .
\end{equation}
Here $W^\kappa(\chi)$ is $\kappa$ kernel defined by Eq.\,(\ref{eq:Wk}), $\delta_m$ is the density fluctuations, $\Omega_m$ is the total matter density today, $H_0$ is the Hubble constant today, and $a \equiv 1/(1 + z)$ is the scale factor.
Galaxy overdensity and CMB lensing convergence $\kappa$ are both projections of 3D density fields, expressed as line-of-sight integrals over their respective projection kernels. The angular cross-correlation power spectrum, adopting the Limber approximation \cite{Limber:1954zz}, is 
\begin{equation}
    C_\ell^{\kappa g} =\int d\chi W^\kappa(\chi)W^g(\chi)\frac{1}{\chi^2}P_{mg}\left(k = \frac{\ell+1/2}{\chi};z\right) \ .
\label{eqn:ckg}
\end{equation}

The Limber approximation is inaccurate for $\ell<10$, but such very large-scale modes are excluded from our analysis, due to poor SNR. The above expression assumes spatial flatness. Here $W^\kappa$ and $W^g$ are the projection kernels for $\kappa$ and the group number density fields
\begin{equation}\label{eq:Wg}
    W^g(z) = n(z) = \frac{c}{H(z)}W^g(\chi) \ .
\end{equation}
\begin{equation}\label{eq:Wk}
    W^\kappa(z) = \frac{3}{2c}\Omega_{m0}\frac{H_0^2}{H(z)}(1+z)\frac{\chi(\chi_*-\chi)}{\chi_*} = \frac{c}{H(z)}W^\kappa(\chi) \ .
\end{equation}
Here $\chi$ is the comoving distance to redshift $z$ and $\chi_* = \chi(z_*\approx1089)$ is the distance to the surface of the last scattering. $n(z)$ is the normalized redshift distribution of galaxy groups.  
$P_{\rm mg}$ is the 3D cross spectrum between matter and group number overdensity.  We define the galaxy bias through $b_{\rm g}\equiv P_{\rm mg}/P_{\rm mm}$. $b_{\rm g}$ is approximately scale-independent at the large scales of interest. The galaxy bias is only redshift-dependent in our analysis, and the bias distribution of four galaxy populations is given by Ref \cite{DESI:2016fyo}.

With the galaxy overdensity map and $\kappa$ map, we use \texttt{HEALPix} to measure the spherical harmonic coefficients $\delta_{\ell m}$ and $\kappa_{\ell m}$. All maps presented in the galaxy-$\kappa$ cross-correlation use {\fontfamily{lmtt}\selectfont
Nside} = 512 and all relevant spherical harmonic transforms use $\ell_{\rm max} = 1024$. 
For simplicity, we adopt the binary mask of AliCPT ``deep patch'' in this calculation. 
The noise in the reconstructed CMB lensing maps dominates the lensing potential signal. In order to enhance the correlation signals, we Weiner-filter the $\kappa$ maps. Hence, the resulting cross-power spectrum reads
\begin{equation} 
\label{Eq Cell sum}
    C^{\kappa g,{\rm WF}}_{\ell} = \frac{1}{2\ell+1}\sum\limits_{m=-\ell}^\ell \delta_{\ell m}\kappa_{\ell m}^*\times\left(\frac{C^{\kappa\kappa}_{\ell}}{C^{\kappa\kappa}_{\ell}+N^{\kappa\kappa}_{\ell}}\right)  \ .
\end{equation}

For the galaxy mocks, we follow the DESI consortium. 
The DESI science parper \cite{DESI:2016fyo} provides the redshift distribution and bias distribution for the bright galaxies (BGS), luminous red galaxies (LRGs), emission line galaxies (ELGs), and quasi-stellar objects (QSOs). The redshift range of BGS, LRG, ELG, and QSO targets will cover $0.05 < z < 0.4$, $0.4 < z < 1.0$, $0.6 < z < 1.6$ and $z < 2.1$, respectively. For QSOs, they use the samples as direct tracers of dark matter in the redshift range $0.9 < z < 2.1$, but not including the foreground neutral-hydrogen absorption systems that make up the Ly-$\alpha$ forest at higher redshifts. The characteristics of their baseline samples for each of these target classes are summarized in Table 3.1 in Ref \cite{DESI:2016fyo}. For LRGs, ELGs, and QSOs, they assume a value for the ratio of galaxy clustering to dark matter clustering, commonly refered to as the large-scale structure bias. On large scales this may be approximated as a function of redshift which is scale-independent, $b(z)$. Except BGS, DESI consortium assumes fiducial biases follow constant $b(z)D(z)$, where $D(z)$ is the linear growth factor normalized by $D(z=0)\equiv 1$. They assume a bias of the form $b_{\rm LRG}(z)D(z) = 1.7$ for LRGs, $b_{\rm ELG}(z)D(z) = 0.84$ for ELGs, and $b_{\rm QSO}(z)D(z) = 1.2$ for quasars. For BGS, the bias is given in the numerical form \cite{DESI:2016fyo}. In the second redshift bin, the redshift of LRGs and ELGs are overlapping. Therefore, the effective bias $\Bar{b} = f_{\rm LRG}b_{\rm LRG} + f_{\rm ELG}b_{\rm ELG}$, where $f_{\rm LRG}$, $f_{\rm ELG}$ is the galaxy size fraction of LRGs and ELGs at $0.4 < z < 1.0$, respectively. The increases in the bias with redshift for various galaxy types are in agreement with observations. These forms keep the observed clustering amplitude of each individual tracer constant with redshift.

\begin{figure}
    \centering
    \includegraphics[width=.8 \textwidth]{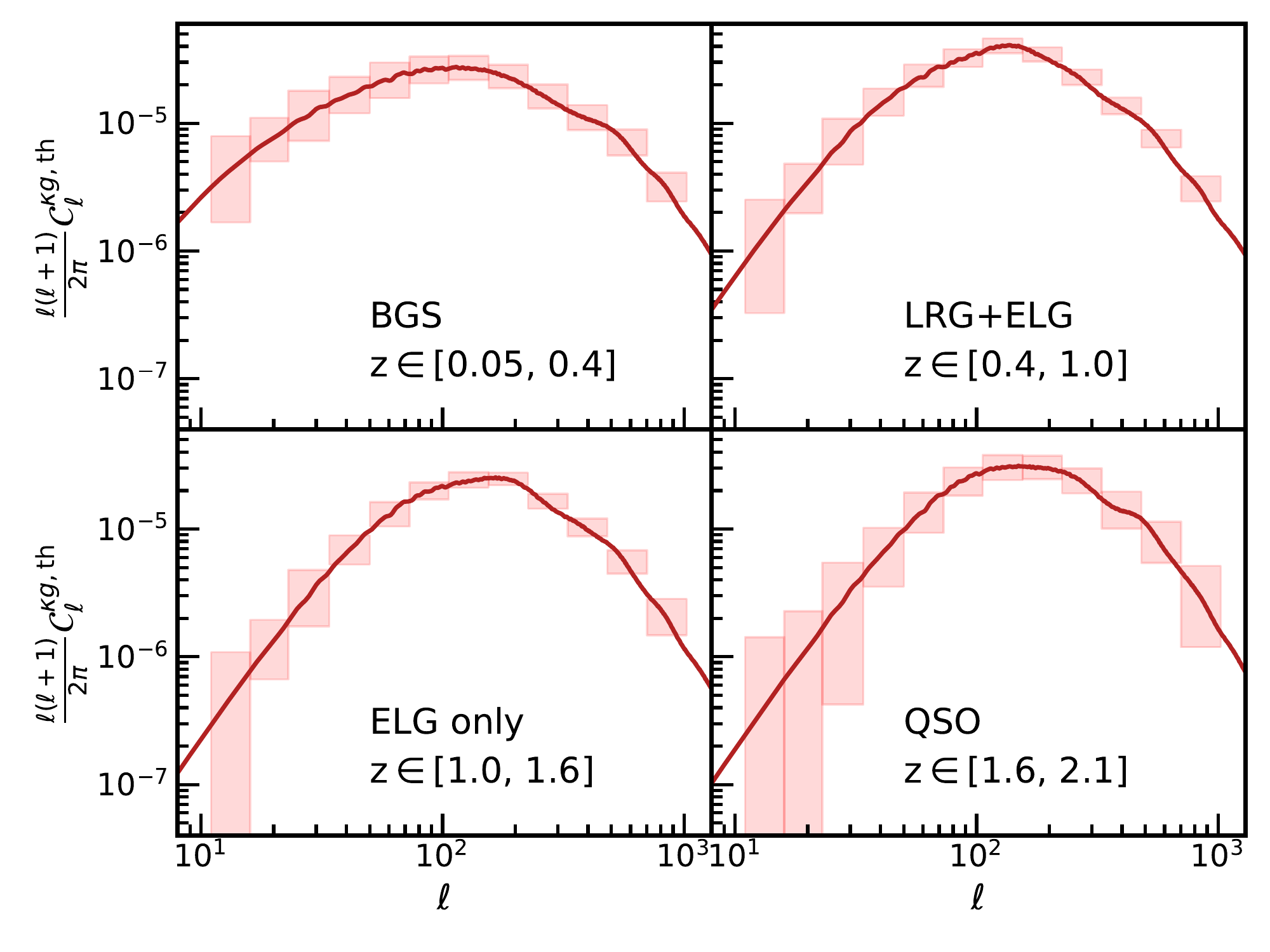}
    \caption{Theoretical cross spectra with DESI target galaxies (BGS, LRG, ELG, and QSO) in four redshift bins. Note that all theoretical calculations include Wiener filter for CMB lensing and binary survey mask in the cross spectra, with the former suppressing the power mainly at small scales, and the latter suppressing the power at all scales. The errors come from the 300 galaxy simulated maps. The four redshift bins achieve $10-20\sigma$ detections of non-zero cross-correlation signal. We divide the $\ell$ bins from $\ell=10$ to 1024 in logarithmic scale.}
    \label{fig:Clkg}
\end{figure}

The evaluation of the theoretical models involves computation of the linear theory power spectrum. We use the Core Cosmology Library \cite{LSSTDarkEnergyScience:2018yem} (\texttt{CCL}\footnote{\url{https://github.com/LSSTDESC/CCL}}) for the theoretical spectrum calculations, with a fiducial Planck 2018 cosmology \cite{Planck:2018vyg}: $\Omega_m = 0.315, \Omega_\Lambda = 0.685, n_s = 0.965, h = H_0/(100 \ \rm km \ s^{-1} \ Mpc^{-1}) = 0.674$ and $\sigma_8 = 0.811$. Given the redshift distribution, bias distribution, and baseline sample sizes of four DESI galaxy types, we can calculate the galaxy auto-power spectra $C_\ell^{\rm gg}$ and white noise spectra $N_\ell^{\rm gg} \equiv 4\pi f_{\rm sky} / N_{\rm gal}$ for each redshift bin. We adopt the DESI baseline survey configuration, which plans to cover 14,000 $\rm deg^2$. We generate 300 galaxy mocks from the total galaxy power spectra $C_\ell^{\rm gg} + N_\ell^{\rm gg}$ by function {\fontfamily{lmtt}\selectfont
synfast} of \texttt{HEALPix}. The CMB lensing mocks are constructed from 300 simulations of the $150$ GHz MV estiamtor assuming ``$\texttt{4 module*yr}$'' stage configuration. As the reference, the corresponding spectrum can be found from the right panel of Figure \ref{fig:recon_cl_4mod}.  

We measure the uncertainty of cross spectra from the 300 simulated galaxy maps and 300 simulated AliCPT-1 CMB lenisng reconstructions. The covariance matrix reads
\begin{equation}
\label{eq:covariance}
    \textbf{Cov}_{\ell\ell^{\prime}} = \frac {1}{N-1} \sum\limits_{n=1}^{N=300} \bigg[\Big(C_{n}^{\kappa g,{\rm WF}}(\ell) - \overline{C}^{\kappa g,{\rm WF}}(\ell)\Big)\times \Big(C_{n}^{\kappa g,{\rm WF}}(\ell^{\prime}) - \overline{C}^{\kappa g,{\rm WF}}(\ell^{\prime})\Big) \bigg] \ .
\end{equation}
Here $C_{n}^{\kappa g,{\rm WF}}(\ell)$ are calculated according to Eq. (\ref{Eq Cell sum}) and $\overline{C}^{\kappa g,{\rm WF}}(\ell)$ is the average cross spectrum. The lensing map is reconstruction noise dominated. The galaxy clustering-CMB lensing cross-correlation coefficient is much smaller than unity, due to mismatch in their redshift distribution. For the two reasons, cosmic variance arising from the galaxy-CMB lensing cross-correlation signal is negligible in the covariance matrix. Fig.\,\ref{fig:Clkg} show the theoretical cross spectra in four redshift bins. The LRGs and ELGs are overlapping in the redshift range $0.4 < z < 1.0$. Remarkably, the theoretical cross spectra are modeled with a binary survey mask and Wiener filter for CMB lensing, that's why the shape of power spectra are not like the conventional one. 

The detections are expected to be significant at all redshift bins. We quantify the detection significance SNR, which is data-driven, and describes the detection significance of a non-zero signal. Here 
\begin{equation}\label{eq:chi2_null}
    S/N \equiv \sqrt{\chi^2_{\rm null}} \ , \quad \chi^2_{\rm null} = \sum\limits_{\ell\ell^{\prime}} C_{\ell}^{\kappa g,\rm th} \textbf{Cov}_{\ell\ell^{\prime}}^{-1}  C_{\ell^{\prime}}^{\kappa g,\rm th} \ . 
\end{equation}
The four redshift bins achieve SNR $=10 - 20$ (Table.\,\ref{tab:SNR}). Since the covariance between different redshift bins is negligible, the total SNR combining all redshift bins is
\begin{equation} \label{eq:SNR}
    \bigg(\frac{S}{N}\bigg)_{\rm total} = \sqrt{ \sum\limits_{\beta} \bigg(\frac{S}{N}\bigg)_{\beta}^{2} } \ .
\end{equation}
Here $\beta=1,\cdots 4$ denotes the for redshift bins. The total SNR is $31.9$. 

\begin{table}
    \centering
    \begin{tabular}{cccccccc} 
    	\hline
    	$z$ & targets & S/N\\
    	\hline 
    	0.05-0.4 & BGS & 13.4\\
        0.4-1.0 & LRG + ELG & 18.7\\
        1.0-1.6 & ELG only & 19.4\\
        1.6-2.1 & QSO & 10.6\\
    	\hline
    \end{tabular}
    \caption{The DESI targets in four redshift bins and their detection significance of non-zero cross-correlation signal. The total SNR is $31.9$.}
    \label{tab:SNR}
\end{table}

The measured cross-correlation has rich cosmological applications, such as constraining $\sigma_8(z)$ and dark energy equation of state. Here we demonstrate its power in testing general relativity. The cross-correlation, in combination with the galaxy auto-correlation, allows us to measure the linear growth rate $D(z)$ ($\propto C_{g\kappa} / \sqrt{C_{gg}}$). For brevity, we fix the shape of matter power spectrum, then
\begin{equation}
    \frac{\sigma_D}{D} \approx \Big( \frac{S}{N} \Big) ^{-1} \ .
\end{equation}
where $\sigma_D/D$, the relative error of $D$, is approximately equal to the inverse of the SNR at each redshift interval that list in Table.\,\ref{tab:SNR}.

In General Relativity (GR), $D(z)$ is completely specified by the expansion history even in the presence of dark energy. The linear growth rate, $f(a)$, is related to the linear growth function $D(a)$, and in GR is given by a good approximation of $\Omega_m(z)$, 
\begin{equation}
    f \equiv \frac{d\ln D}{d\ln a} = \frac{a}{D}\frac{dD}{da} \simeq \Omega_m^\gamma(z) \ .
\end{equation}
where $\gamma$ is the growth index, approximately equal to $0.55$ in GR. $\Omega_m(z)$ is the fraction of the total matter density at redshift $z$. In the alternative gravity theories, a widely adopted parameterization of the modified growth rate is to alter the growth index $\gamma$. The uncertainty of $\gamma$ is given by
\begin{equation}
    \sigma_\gamma^2 = \Big[ \sum\limits_z \frac{(\partial\ln D / \partial\ln\gamma)^2}{(\sigma_D / D)^2} \Big]^{-1} \ .
\end{equation}

We can constrain $\gamma$ to $\sigma_\gamma$ = 0.16, which is approximately $29\%$ relative error in the growth index determination. This constraint will be complementary to other constraints (e.g. the DESI forecast $\sigma_\gamma = 0.04$ \cite{Stril:2010}). 

All the above results are obtained by assuming the ``$\texttt{4 module*yr}$'' stage configuration. For the ``$\texttt{48 module*yr}$'' stage configuration, the galaxy-CMB lensing cross-correlation SNR reaches $53$, and the $\sigma_\gamma = 0.10$, namely the relative error in the growth index is about $18\%$. In this analysis, we have neglected the cosmic variance caused by the cross-correlation signal. We checked that, including the cosmic variance only reduces the signal-to-noise slightly (e.g. 53.1 to 52.2 for ``$\texttt{48 module*yr}$'').

\subsection{Cross-correlations with Cosmic Infrared Background}
\label{sec:XCIB}

The cosmic infrared background is the far-infrared relic emission of the galaxies during their formation and evolution processes. 
Produced by the heated gas within the galaxies, the CIB mainly consists of the integrated emission from unresolved dusty star-forming galaxies (DSFGs). Therefore, it contains a wealth of information about the the dusty star-forming galaxies distribution at high redshift. And with the extraordinary redshift depth of CIB observation, the CIB anisotropies is thus an excellent tool to trace the underlying dark matter halos in which the galaxies reside and to probe the connection between luminous matter and dark matter.

In Refs \cite{Planck:2013oqw,Planck:2013qqi}, a strong correlation (about $80\%$) is observed between the CIB anisotropies and a lensing-derived projected mass map. Therefore, we can write the CIB-lensing cross-spectra as
\begin{equation}
C_\ell^{\phi \nu}=\int_0^{\chi_*} d\chi\frac{1}{\chi^2}W^{\nu}W^{\phi}P_{mg}(k=\ell/\chi,z)\;,
\end{equation}
where $\phi$ is the lesing potential and $\nu$ represent CIB intensity at frequency $\nu$; the integral is over $\chi$, the comoving distance along the line of sight; $\chi_*$ is the comoving distance to the last scattering surface, $P_{mg}(k=\ell/\chi,z)$ is the cross-correlation between dark matter and dusty galaxies using Limber approximation. $W^{\nu}$ and $W^{\phi}$ are the redshift weights for the CIB and lesning potential, respectively
\begin{equation}
\begin{aligned}
W^{\nu}
&=aj(\nu,\chi)\;,\\
W^{\phi}\,
&=-3\frac{\Omega_m}{a}\frac{H_0^2}{c^2k^2} \left (\frac{\chi_*-\chi}{\chi \chi_*}\right)\;,\\
&=-3\frac{\Omega_m}{a}\frac{H_0^2\chi^2}{c^2\ell^2} \left (\frac{\chi_*-\chi}{\chi \chi_*}\right)\;.
\end{aligned}
\end{equation}
Here, $j(\nu,\chi)$ is the CIB emissivity at frequency $\nu$, $a$ is the scale factor, $H_0$ is the Hubble parameter today, $\Omega_m$ is the matter density today in critical density unit.
To make the result consistent with both CIB auto and CIB-lensing cross-spectra of Planck multi-frequency measurements, we adopted the halo occupation distribution (HOD) model in Planck paper \cite{Planck:2011ivn,Planck:2013qqi} to calculate $P_{mg}$. The details can be found in Ref. \cite{Song:2002sg}. 

We also perform the error estimation of the CIB-lensing potential cross-power spectrum from AliCPT and Planck. Considering only the Gaussian statistical errors, we calculate SNR using a simple Fisher matrix prescription~\cite{Planck:2013qqi}

\begin{table}
\centering
\caption{Lensing-CIB cross-correlation SNRs}
\label{tab:cibsnr}
\vspace{2mm}
\begin{tabular}[h]{|c|c|c|} 
    \hline
     ${\rm frequency(GHz)}$ & $\texttt{4 module*yr}$ & $\texttt{48 module*yr}$  \\
    \hline
     353& 18.2 & 25.1\\

     545& 19.3 & 33.2\\

     857& 23.1 & 42.2\\
    \hline          
    {\rm total}& 23.3 & 43.1\\
    \hline
\end{tabular}
\end{table}

\begin{figure}
    \centering
    \includegraphics[width=1.10 \textwidth]{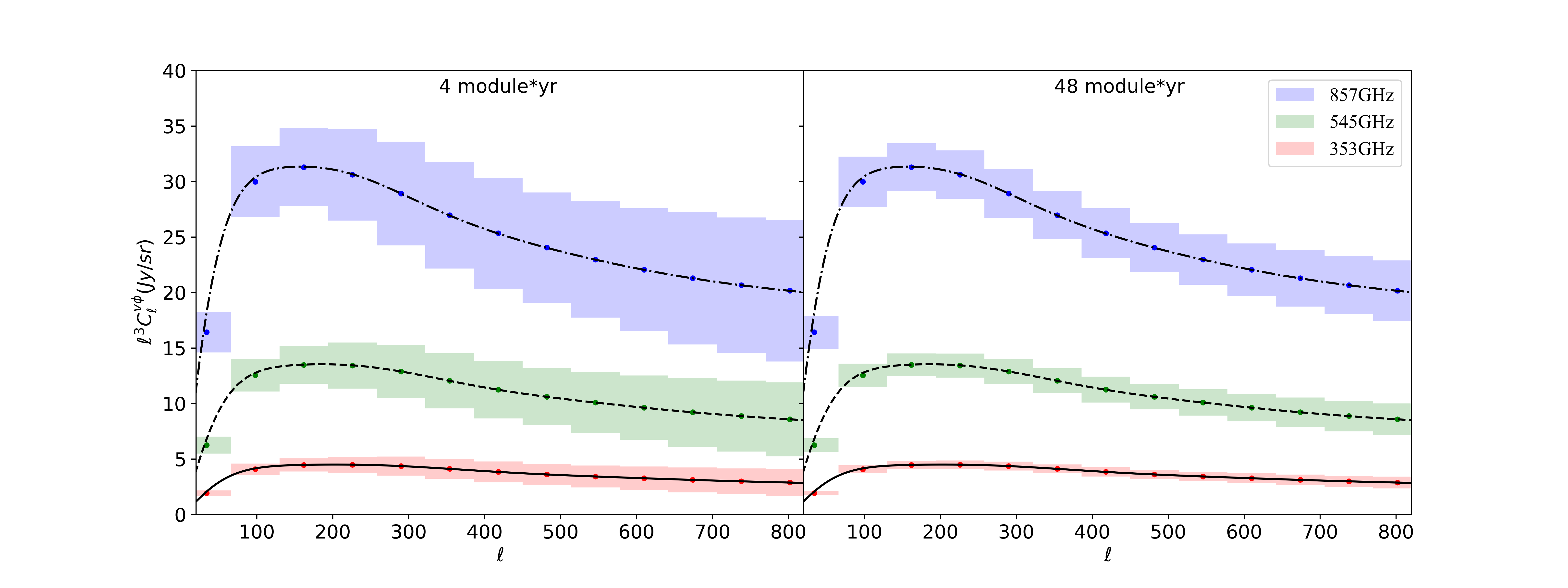}
    \caption{The forecasted Lensing-CIB cross-correlation. In ``$\texttt{4 module*yr}$'' configuration, the SNR are: $18.2,19.3,23.1$ for Planck $353,545,857$ GHz. In ``$\texttt{48 module*yr}$'' configuration, the SNR are: $25.1,33.2,42.2$ for Planck $353,545,857$ GHz.}
    \label{fig:cibxc}
\end{figure}

\begin{equation}
\left(\frac{S}{N}\right)=\sqrt{\sum_{\ell}\sum_{\nu\nu^{\prime}}C_\ell^{\phi \nu}C_\ell^{\phi \nu^{\prime}}{\rm CoV}^{-1}_{\ell}(\nu,\nu^{\prime})}\\
\label{eq:cibsnr}
\end{equation}

\begin{equation}
{\rm CoV}_{\ell}(\nu,\nu^{\prime})=\left\{
\begin{aligned}
&\frac{(C_\ell^{\phi \nu})^2+(C_\ell^{\nu\nu}+N_\ell^{\nu\nu})(C_\ell^{\phi\phi}+N_{\ell}^{\phi\phi})}{(2\ell+1)f_{sky}\Delta \ell}\;,\;(\nu=\nu^{\prime})\\
&\\
&\frac{(C_\ell^{\phi \nu})(C_l^{\phi \nu^{\prime}})+(C_\ell^{\nu\nu^{\prime}}+N_\ell^{\nu\nu^{\prime}})(C_\ell^{\phi\phi}+N_{\ell}^{\phi\phi})}{(2\ell+1)f_{sky}\Delta \ell}\;,\;(\nu \neq \nu^{\prime})
\end{aligned}
\right.
\label{eq:cibcov}
\end{equation}
Here, $\Delta \ell$ is the width of the ell bins; $C_\ell^{\nu\nu}$ and $C_\ell^{\phi\phi}$ are the auto-power spectrum of CIB intensity and lensing potential, respectively; while the corresponding noise spectra are written as $N_\ell^{\nu\nu}$ and $N_\ell^{\phi\phi}$.
Furthermore, $C_\ell^{\nu\nu'}$ and $N_\ell^{\nu\nu'}$ are the measured CIB intensity cross-correlation signal and noise spectra between different frequencies. 
The observed data $(C_\ell^{\nu\nu'}, N_\ell^{\nu\nu'}, N_\ell^{\nu\nu})$ are taken from the published Planck Legacy Archive\footnote{\url{https://lambda.gsfc.nasa.gov/product/planck/curr/planck_tp_lenz_get.cfm}}\cite{Lenz:2019ugy}; and to maintain consistency, we apply a binning strategy similar to the one used in \cite{Planck:2013qqi}, namely $\Delta \ell=64$. $N^{\phi\phi}_{\ell}$ are taken the $150$ GHz channel MV-$N0$ noise spectrum.

In Figure \ref{fig:cibxc}, we show the forecasted Lensing-CIB cross spectra. The purple, green and red shaded boxes denote for the 1$\sigma$ scatters for Planck 857, 545 and 353 GHz channel. Black curves denote for the theoretical predictions by assuming the Planck cosmology. The Fisher matrix based SNR are $18.2,19.3,23.1$ for $353,545,857$ GHz in the ``$\texttt{4 module*yr}$'' stage. Under the ``$\texttt{48 module*yr}$'' scenario, the SNR are $25.1,33.2,42.2$ for $353,545$ and $857$ GHz. 
Following the Eq. (\ref{eq:cibsnr}) and (\ref{eq:cibcov}), one can calculate the total SNR.The corresponding results are listed in the last row of Table. \ref{tab:cibsnr}. One can see that due to the strong correlations among different frequencies in the covariance, the combined SNRs are enhanced in a very limited manner compared the 857 GHz channel, which is the highest SNR channel. The final total cross-correlation SNR=$23.3$ and $43.1$ for the AliCPT-1 first and final stage, respectively. We summarize the Lensing-CIB cross-correlation SNRs in Table \ref{tab:cibsnr}. 

\section{Conclusion}
\label{sec:con}

AliCPT-1 project is the first Chinese CMB experiment aiming for high precision measurement of CMB polarization. The experiment observes at $90$ and $150$ GHz frequencies with an intermediate spatial resolution (FWHM=$19,11$ arcmin for $90$ and $150$ GHz, respectively) and a 3rd generation CMB experiment noise level.  
The harmonic means of the noise variance in the ``$\texttt{4 module*yr}$'' case are about $11~\mu{\rm K}\cdot{\rm arcmin}$ at $90$ GHz and $17~\mu{\rm K}\cdot{\rm arcmin}$ at $150$ GHz.

In this paper, we investigate the ability of AliCPT-1 for measuring the CMB lensing signal. We forecast the lensing reconstruction, lensing-galaxy cross-correlation as well as the lensing-CIB cross-correlation in AliCPT-1. In details, we consider two mission designs, namely the first (``$\texttt{4 module*yr}$'') and final (``$\texttt{48 module*yr}$'') stages. We adopt the technically mature quadratic estimator for the lensing reconstruction. We find that, in the first stage, the SNR for T-only estimator are $3.5$ and $8.3$ in $90$ and $150$ GHz channel, respectively. For P-only estimator, the SNR are $5.1$ for $90$ GHz and $6.6$ for $150$ GHz. For MV estimator, the SNR are $9.2$ and $15.4$ for $90$ and $150$ GHz channel, respectively. In the final stage, for T-only estimator the SNR are $4.8$ and $8.0$ in $90$ and $150$ GHz channel, respectively.
For P-only estimator, the SNR are $21.7$ for $90$ GHz and $25.4$ for $150$ GHz. For MV estimator, the SNR are $24.3$ and $31.1$ for $90$ and $150$ GHz channel, respectively. Unlike Planck data, the polarization data in AliCPT-1 plays an essential role in the lensing reconstruction due to the excellent noise level in polarization pattern. We find that the EB estimator will dominate the lensing reconstruction once we accumulate the data and arrive at the final stage. 
In this work, we focus on the statistical noise according to the ``deep patch'' scanning strategy. We do not consider the systematic effects and foreground residual contamination to the lensing reconstruction. We leave these effects for the future studies.   

For lensing-galaxy cross-correlation, we cross-correlate the AliCPT-1 lensing (150 GHz MV estimator) with DESI bright galaxy samples (BGS), luminous red galaxies (LRGs), emission line galaxies (ELGs), and quasi-stellar objects (QSOs). In the first stage, we report the cross-correlation SNR $=10-20$ for the redshift bins arranged from $z=0.05$ to $2.1$.
The total SNR can reach $32$. Furthermore, we show the constraint on the growth index $\gamma$ as an example for the cosmology application of these cross-correlation signals. We find $\sigma_\gamma=0.16$, which is approximately $29\%$ relative error in the growth index determination. 
In the final stage, the total cross-correlation SNR can reach $52$ and the growth index determination error can be further reduced to the $\sigma_\gamma=0.10$ level.

For lensing-CIB cross-correlation, we forecast the AliCPT-1 lensing (150 GHz MV estimator) cross-correlate with Planck CIB from 353, 545 and 857 GHz channels. In the first stage, we report SNR=$18.2$ for $353$ GHz, SNR=$19.3$ for $545$ GHz and SNR=$23.1$ for $857$ GHz. In the final stage, these three numbers are SNR=$25.1,33.2$ and $42.2$.
The final total lensing-CIB cross-correlation by combining the three frequencies in Planck CIBs are SNR=$23.3$ and $43.1$ for the AliCPT-1 first and final stage, respectively.

Thanks to the excellent detector noise performance, the AliCPT-1 mission can measure the lensing signal with a fairly good significance, especially via the polarization data. We believe, we present a comprehensive studies on the ability of AliCPT-1 for measuring lensing signal as well as the relevant cosmology applications.   

\acknowledgments
We acknowledge Zuhui Fan for various comments on the preliminary version of the draft. This work are supported by the National Key R\&D Program of China (No. 2020YFC2201603, No.2020YFC2201601, No. 2020YFC2201600), National Natural Science Foundation of China (No. 11653003), the 111 project (No. B20019).

\newpage



\begin{thebibliography}{99}

\bibitem[Blanchard, Schneider(1987)]{1987A&A...184....1B} Blanchard, A. and Schneider, J.\ 1987, 184, 1

\bibitem{Kamionkowski:1996ks}
M.~Kamionkowski, A.~Kosowsky and A.~Stebbins,
Phys. Rev. D \textbf{55}, 7368-7388 (1997)
doi:10.1103/PhysRevD.55.7368
[arXiv:astro-ph/9611125 [astro-ph]].

\bibitem{Kamionkowski:1996zd}
M.~Kamionkowski, A.~Kosowsky and A.~Stebbins,
Phys. Rev. Lett. \textbf{78}, 2058-2061 (1997)
doi:10.1103/PhysRevLett.78.2058
[arXiv:astro-ph/9609132 [astro-ph]].

\bibitem{Zaldarriaga:1998ar}
M.~Zaldarriaga and U.~Seljak,
Phys. Rev. D \textbf{58}, 023003 (1998)
doi:10.1103/PhysRevD.58.023003
[arXiv:astro-ph/9803150 [astro-ph]].

\bibitem{Ade:2013tyw}
P.~A.~R.~Ade \textit{et al.} [Planck],
Astron. Astrophys. \textbf{571}, A17 (2014)
doi:10.1051/0004-6361/201321543
[arXiv:1303.5077 [astro-ph.CO]].

\bibitem{Ade:2015zua}
P.~A.~R.~Ade \textit{et al.} [Planck],
Astron. Astrophys. \textbf{594}, A15 (2016)
doi:10.1051/0004-6361/201525941
[arXiv:1502.01591 [astro-ph.CO]].

\bibitem{Aghanim:2018oex}
N.~Aghanim \textit{et al.} [Planck],
Astron. Astrophys. \textbf{641}, A8 (2020)
doi:10.1051/0004-6361/201833886
[arXiv:1807.06210 [astro-ph.CO]].

\bibitem{Lewis:2006fu}
A.~Lewis and A.~Challinor,
Phys. Rept. \textbf{429}, 1-65 (2006)
doi:10.1016/j.physrep.2006.03.002
[arXiv:astro-ph/0601594 [astro-ph]].

\bibitem{Hanson:2009kr}
D.~Hanson, A.~Challinor and A.~Lewis,
Gen. Rel. Grav. \textbf{42}, 2197-2218 (2010)
doi:10.1007/s10714-010-1036-y
[arXiv:0911.0612 [astro-ph.CO]].

\bibitem{Hu:2001kj}
W.~Hu and T.~Okamoto,
Astrophys. J. \textbf{574}, 566-574 (2002)
doi:10.1086/341110
[arXiv:astro-ph/0111606 [astro-ph]].

\bibitem{Okamoto:2003zw}
T.~Okamoto and W.~Hu,
Phys. Rev. D \textbf{67}, 083002 (2003)
doi:10.1103/PhysRevD.67.083002
[arXiv:astro-ph/0301031 [astro-ph]].

\bibitem{vanEngelen:2012va}
A.~van Engelen, R.~Keisler, O.~Zahn, K.~A.~Aird, B.~A.~Benson, L.~E.~Bleem, J.~E.~Carlstrom, C.~L.~Chang, H.~M.~Cho and T.~M.~Crawford, \textit{et al.}
Astrophys. J. \textbf{756}, 142 (2012)
doi:10.1088/0004-637X/756/2/142
[arXiv:1202.0546 [astro-ph.CO]].

\bibitem{Hanson:2013hsb}
D.~Hanson \textit{et al.} [SPTpol],
Phys. Rev. Lett. \textbf{111}, no.14, 141301 (2013)
doi:10.1103/PhysRevLett.111.141301
[arXiv:1307.5830 [astro-ph.CO]].

\bibitem{Sayre:2019dic}
J.~T.~Sayre \textit{et al.} [SPT],
Phys. Rev. D \textbf{101}, no.12, 122003 (2020)
doi:10.1103/PhysRevD.101.122003
[arXiv:1910.05748 [astro-ph.CO]].

\bibitem{Wu:2019hek}
W.~L.~K.~Wu, L.~M.~Mocanu, P.~A.~R.~Ade, A.~J.~Anderson, J.~E.~Austermann, J.~S.~Avva, J.~A.~Beall, A.~N.~Bender, B.~A.~Benson and F.~Bianchini, \textit{et al.}
Astrophys. J. \textbf{884}, 70 (2019)
doi:10.3847/1538-4357/ab4186
[arXiv:1905.05777 [astro-ph.CO]].


\bibitem{Hirata:2004rp}
C.~M.~Hirata, N.~Padmanabhan, U.~Seljak, D.~Schlegel and J.~Brinkmann,
Phys. Rev. D \textbf{70}, 103501 (2004)
doi:10.1103/PhysRevD.70.103501
[arXiv:astro-ph/0406004 [astro-ph]].

\bibitem{Smith:2007rg}
K.~M.~Smith, O.~Zahn and O.~Dore,
Phys. Rev. D \textbf{76}, 043510 (2007)
doi:10.1103/PhysRevD.76.043510
[arXiv:0705.3980 [astro-ph]].

\bibitem{Hirata:2008cb}
C.~M.~Hirata, S.~Ho, N.~Padmanabhan, U.~Seljak and N.~A.~Bahcall,
Phys. Rev. D \textbf{78}, 043520 (2008)
doi:10.1103/PhysRevD.78.043520
[arXiv:0801.0644 [astro-ph]].

\bibitem{Feng:2012uf}
C.~Feng, G.~Aslanyan, A.~V.~Manohar, B.~Keating, H.~P.~Paar and O.~Zahn,
Phys. Rev. D \textbf{86}, 063519 (2012)
doi:10.1103/PhysRevD.86.063519
[arXiv:1207.3326 [astro-ph.CO]].

\bibitem{Smidt:2010by}
J.~Smidt, A.~Cooray, A.~Amblard, S.~Joudaki, D.~Munshi, M.~G.~Santos and P.~Serra,
Astrophys. J. Lett. \textbf{728}, L1 (2011)
doi:10.1088/2041-8205/728/1/L1
[arXiv:1012.1600 [astro-ph.CO]].

\bibitem{Feng:2011jx}
C.~Feng, B.~Keating, H.~P.~Paar and O.~Zahn,
Phys. Rev. D \textbf{85}, 043513 (2012)
doi:10.1103/PhysRevD.85.043513
[arXiv:1111.2371 [astro-ph.CO]].

\bibitem{Ade:2013gez}
P.~A.~R.~Ade \textit{et al.} [POLARBEAR],
Phys. Rev. Lett. \textbf{113}, 021301 (2014)
doi:10.1103/PhysRevLett.113.021301
[arXiv:1312.6646 [astro-ph.CO]].

\bibitem{Ade:2013hjl}
P.~A.~R.~Ade \textit{et al.} [POLARBEAR],
Phys. Rev. Lett. \textbf{112}, 131302 (2014)
doi:10.1103/PhysRevLett.112.131302
[arXiv:1312.6645 [astro-ph.CO]].

\bibitem{Das:2011ak}
S.~Das, B.~D.~Sherwin, P.~Aguirre, J.~W.~Appel, J.~R.~Bond, C.~S.~Carvalho, M.~J.~Devlin, J.~Dunkley, R.~Dunner and T.~Essinger-Hileman, \textit{et al.}
Phys. Rev. Lett. \textbf{107}, 021301 (2011)
doi:10.1103/PhysRevLett.107.021301
[arXiv:1103.2124 [astro-ph.CO]].

\bibitem{Das:2013zf}
S.~Das, T.~Louis, M.~R.~Nolta, G.~E.~Addison, E.~S.~Battistelli, J.~R.~Bond, E.~Calabrese, D.~C.~M.~J.~Devlin, S.~Dicker and J.~Dunkley, \textit{et al.}
JCAP \textbf{04}, 014 (2014)
doi:10.1088/1475-7516/2014/04/014
[arXiv:1301.1037 [astro-ph.CO]].

\bibitem{Array:2016afx}
P.~A.~R.~Ade \textit{et al.} [BICEP2 and Keck Array],
Astrophys. J. \textbf{833}, no.2, 228 (2016)
doi:10.3847/1538-4357/833/2/228
[arXiv:1606.01968 [astro-ph.CO]].

\bibitem{Planck:2013qqi}
P.~A.~R.~Ade \textit{et al.} [Planck],
Astron. Astrophys. \textbf{571}, A18 (2014)
doi:10.1051/0004-6361/201321540
[arXiv:1303.5078 [astro-ph.CO]].

\bibitem{Lewis:1999bs}
A.~Lewis, A.~Challinor and A.~Lasenby,
Astrophys. J. \textbf{538}, 473-476 (2000)
doi:10.1086/309179
[arXiv:astro-ph/9911177 [astro-ph]].

\bibitem{Gorski:2004by}
K.~M.~Gorski, E.~Hivon, A.~J.~Banday, B.~D.~Wandelt, F.~K.~Hansen, M.~Reinecke and M.~Bartelman,
Astrophys. J. \textbf{622}, 759-771 (2005)
doi:10.1086/427976
[arXiv:astro-ph/0409513 [astro-ph]].

\bibitem[Carron(2020)]{2020ascl.soft10010C} Carron, J.\ 2020, Astrophysics Source Code Library. ascl:2010.010

\bibitem{Carron:2017mqf}
J.~Carron and A.~Lewis,
Phys. Rev. D \textbf{96}, no.6, 063510 (2017)
doi:10.1103/PhysRevD.96.063510
[arXiv:1704.08230 [astro-ph.CO]].

\bibitem{Kesden:2003cc}
M.~H.~Kesden, A.~Cooray and M.~Kamionkowski,
Phys. Rev. D \textbf{67}, 123507 (2003)
doi:10.1103/PhysRevD.67.123507
[arXiv:astro-ph/0302536 [astro-ph]].

\bibitem{Planck:2013oqw}
P.~A.~R.~Ade \textit{et al.} [Planck],
Astron. Astrophys. \textbf{571}, A1 (2014)
doi:10.1051/0004-6361/201321529
[arXiv:1303.5062 [astro-ph.CO]].

\bibitem{Planck:2013qqi}
P.~A.~R.~Ade \textit{et al.} [Planck],
Astron. Astrophys. \textbf{571}, A18 (2014)
doi:10.1051/0004-6361/201321540
[arXiv:1303.5078 [astro-ph.CO]].

\bibitem{Planck:2011ivn}
P.~A.~R.~Ade \textit{et al.} [Planck],
Astron. Astrophys. \textbf{536}, A18 (2011)
doi:10.1051/0004-6361/201116461
[arXiv:1101.2028 [astro-ph.CO]].

\bibitem{Song:2002sg}
Y.~S.~Song, A.~Cooray, L.~Knox and M.~Zaldarriaga,
Astrophys. J. \textbf{590}, 664-672 (2003)
doi:10.1086/375188
[arXiv:astro-ph/0209001 [astro-ph]].

\bibitem{LSSTDarkEnergyScience:2018yem}
N.~E.~Chisari \textit{et al.} [LSST Dark Energy Science],
Astrophys. J. Suppl. \textbf{242}, no.1, 2 (2019)
doi:10.3847/1538-4365/ab1658
[arXiv:1812.05995 [astro-ph.CO]].

\bibitem{Bethermin:2012ki}
M.~Bethermin, E.~Daddi, G.~Magdis, M.~T.~Sargent, Y.~Hezaveh, D.~Elbaz, D.~L.~Borgne, J.~Mullaney, M.~Pannella and V.~Buat, \textit{et al.}
Astrophys. J. Lett. \textbf{757}, L23 (2012)
doi:10.1088/2041-8205/757/2/L23
[arXiv:1208.6512 [astro-ph.CO]]. 

\bibitem{Lenz:2019ugy}
D.~Lenz, O.~Dor\'e and G.~Lagache,
Astrophys. J. \textbf{883}, no.1, 75 (2019)
doi:10.3847/1538-4357/ab3c2b
[arXiv:1905.00426 [astro-ph.CO]].

\bibitem{Smith:2007rg}
K.~M.~Smith, O.~Zahn and O.~Dore,
Phys. Rev. D \textbf{76}, 043510 (2007)
doi:10.1103/PhysRevD.76.043510
[arXiv:0705.3980 [astro-ph]].

\bibitem{Maniyar:2021msb}
A.~S.~Maniyar, Y.~Ali-Ha\"\i{}moud, J.~Carron, A.~Lewis and M.~S.~Madhavacheril,
Phys. Rev. D \textbf{103}, no.8, 083524 (2021)
doi:10.1103/PhysRevD.103.083524
[arXiv:2101.12193 [astro-ph.CO]].

\bibitem{Hirata:2003ka}
C.~M.~Hirata and U.~Seljak,
Phys. Rev. D \textbf{68}, 083002 (2003)
doi:10.1103/PhysRevD.68.083002
[arXiv:astro-ph/0306354 [astro-ph]].

\bibitem{Darwish:2020fwf}
O.~Darwish, M.~S.~Madhavacheril, B.~D.~Sherwin, S.~Aiola, N.~Battaglia, J.~A.~Beall, D.~T.~Becker, J.~R.~Bond, E.~Calabrese and S.~Choi, \textit{et al.}
Mon. Not. Roy. Astron. Soc. \textbf{500}, no.2, 2250-2263 (2020)
doi:10.1093/mnras/staa3438
[arXiv:2004.01139 [astro-ph.CO]].

\bibitem{Kitanidis:2020xno}
E.~Kitanidis and M.~White,
Mon. Not. Roy. Astron. Soc. \textbf{501}, no.4, 6181-6198 (2021)
doi:10.1093/mnras/staa3927
[arXiv:2010.04698 [astro-ph.CO]].

\bibitem{Marques:2020dsb}
G.~A.~Marques, J.~Liu, K.~M.~Huffenberger and J.~Colin Hill,
Astrophys. J. \textbf{904}, no.2, 182 (2020)
doi:10.3847/1538-4357/abc003
[arXiv:2008.04369 [astro-ph.CO]].

\bibitem{Krolewski:2021yqy}
A.~Krolewski, S.~Ferraro and M.~White,
JCAP \textbf{12}, no.12, 028 (2021)
doi:10.1088/1475-7516/2021/12/028
[arXiv:2105.03421 [astro-ph.CO]].

\bibitem{White:2021yvw}
M.~White, R.~Zhou, J.~DeRose, S.~Ferraro, S.~F.~Chen, N.~Kokron, S.~Bailey, D.~Brooks, J.~Garcia-Bellido and J.~Guy, \textit{et al.}
[arXiv:2111.09898 [astro-ph.CO]].

\bibitem{Geach:2017crt}
J.~E.~Geach and J.~A.~Peacock,
Nature Astron. \textbf{1}, no.11, 795-799 (2017)
doi:10.1038/s41550-017-0259-1
[arXiv:1707.09369 [astro-ph.CO]].

\bibitem{DES:2017fyz}
E.~J.~Baxter \textit{et al.} [DES and SPT],
Mon. Not. Roy. Astron. Soc. \textbf{476}, no.2, 2674-2688 (2018)
doi:10.1093/mnras/sty305
[arXiv:1708.01360 [astro-ph.CO]].

\bibitem{ACT:2020izl}
M.~S.~Madhavacheril \textit{et al.} [ACT],
Astrophys. J. Lett. \textbf{903}, no.1, L13 (2020)
doi:10.3847/2041-8213/abbccb
[arXiv:2009.07772 [astro-ph.CO]].

\bibitem{Sun:2021rhp}
Z.~Sun, J.~Yao, F.~Dong, X.~Yang, L.~Zhang and P.~Zhang,
[arXiv:2109.07387 [astro-ph.CO]].

\bibitem{DESI:2018ymu}
A.~Dey \textit{et al.} [DESI],
Astron. J. \textbf{157}, no.5, 168 (2019)
doi:10.3847/1538-3881/ab089d
[arXiv:1804.08657 [astro-ph.IM]].

\bibitem{DESI:2016fyo}
A.~Aghamousa \textit{et al.} [DESI],
[arXiv:1611.00036 [astro-ph.IM]].

\bibitem{Stril:2010}
A.~Stril, R.~Cahn, and E.~Linder,  
Mon. Not. Roy. Astron. Soc. \textbf{404} 239-246 (2010)
doi:10.1111/j.1365-2966.2010.16193.x
[arXiv:0910.1833 [astro-ph.CO]].

\bibitem{Limber:1954zz}
D.~N.~Limber,
Astrophys. J. \textbf{119}, 655 (1954)
doi:10.1086/145870

\bibitem{LSSTDarkEnergyScience:2018yem}
N.~E.~Chisari \textit{et al.} [LSST Dark Energy Science],
Astrophys. J. Suppl. \textbf{242}, no.1, 2 (2019)
doi:10.3847/1538-4365/ab1658
[arXiv:1812.05995 [astro-ph.CO]].

\bibitem{Planck:2018vyg}
N.~Aghanim \textit{et al.} [Planck],
Astron. Astrophys. \textbf{641}, A6 (2020)
[erratum: Astron. Astrophys. \textbf{652}, C4 (2021)]
doi:10.1051/0004-6361/201833910
[arXiv:1807.06209 [astro-ph.CO]].

\bibitem{Huterer:2013xky}
D.~Huterer, D.~Kirkby, R.~Bean, A.~Connolly, K.~Dawson, S.~Dodelson, A.~Evrard, B.~Jain, M.~Jarvis and E.~Linder, \textit{et al.}
Astropart. Phys. \textbf{63}, 23-41 (2015)
doi:10.1016/j.astropartphys.2014.07.004
[arXiv:1309.5385 [astro-ph.CO]].





\end{thebibliography}
\end{document}